\documentclass[a4paper]{aa}

\usepackage{amsmath}
\usepackage{graphicx}
\usepackage{siunitx}
\usepackage{txfonts}
\usepackage{natbib}
\usepackage{ifthen,hyperref}
\usepackage{fixltx2e}
\usepackage{lscape}
\usepackage{color}

\bibpunct{(}{)}{;}{a}{}{,}

\begin{document}

\title{Estimating extragalactic Faraday rotation\thanks{Posterior samples for the extragalactic contribution to all data points and all results of our fiducial model are provided at \url{http://www.mpa-garching.mpg.de/ift/faraday/}.}}

\author{N.~Oppermann\thanks{\email{niels@cita.utoronto.ca}} \inst{\ref{inst:CITA},\ref{inst:MPA}}
    \and H.~Junklewitz \inst{\ref{inst:MPA},\ref{inst:AIFA}}
    \and M.~Greiner \inst{\ref{inst:MPA}}
    \and T.A.~En{\ss}lin \inst{\ref{inst:MPA},\ref{inst:LMU}}
    \and T.~Akahori \inst{\ref{inst:sydney}}
    \and E.~Carretti \inst{\ref{inst:CSIRO}}
    \and B.M.~Gaensler \inst{\ref{inst:sydney}}
    \and A.~Goobar \inst{\ref{inst:Albanova}}
    \and L.~Harvey-Smith \inst{\ref{inst:CSIRO}}
    \and M.~Johnston-Hollitt \inst{\ref{inst:wellington}}
    \and L.~Pratley \inst{\ref{inst:wellington}}
    \and D.H.F.M.~Schnitzeler \inst{\ref{inst:MPIFR}}
    \and J.M.~Stil \inst{\ref{inst:calgary}}
    \and V.~Vacca \inst{\ref{inst:MPA}}
    }

\institute{Canadian Institute for Theoretical Astrophysics, University of Toronto, 60 St.\ George Street, Toronto ON, M5S 3H8, Canada\label{inst:CITA}
    \and Max Planck Institute for Astrophysics, Karl-Schwarzschild-Str.~1, 85748 Garching, Germany\label{inst:MPA}
    \and Argelander-Institut f\"ur Astronomie, Auf dem H\"ugel 71, 52121 Bonn, Germany\label{inst:AIFA} 
    \and Ludwig-Maximilians-Universit\"at M\"unchen, Fakult\"at f\"ur Physik, D-80799 M\"unchen, Germany\label{inst:LMU}
    \and Sydney Institute for Astronomy, School of Physics, The University of Sydney, NSW 2006, Australia\label{inst:sydney}
    \and CSIRO Astronomy and Space Science, PO Box 76, Epping, NSW 1700, Australia\label{inst:CSIRO}
    \and The Oskar Klein Centre, Physics Department, Stockholm University,
    Albanova University Center, SE 106 91 Stockholm, Sweden\label{inst:Albanova}
    \and School of Chemical \& Physical Sciences, Victoria University of Wellington, PO Box 600, Wellington 6014, New Zealand\label{inst:wellington}
    \and Max Planck Institut f\"ur Radioastronomie, Auf dem H\"ugel 69, 53121 Bonn, Germany\label{inst:MPIFR}
    \and Department of Physics and Astronomy, The University of Calgary, 2500 University Drive NW, Calgary AB, T2N 1N4, Canada\label{inst:calgary}
    }

\date{\today}

\abstract{
Observations of Faraday rotation for extragalactic sources probe magnetic fields both inside and outside the Milky Way. Building on our earlier estimate of the Galactic contribution, we set out to estimate the extragalactic contributions. We discuss the problems involved; in particular, we point out that taking the difference between the observed values and the Galactic foreground reconstruction is not a good estimate for the extragalactic contributions. We point out a degeneracy between the contributions to the observed values due to extragalactic magnetic fields and observational noise and comment on the dangers of over-interpreting an estimate without taking into account its uncertainty information. To overcome these difficulties, we develop an extended reconstruction algorithm based on the assumption that the observational uncertainties are accurately described for a subset of the data, which can overcome the degeneracy with the extragalactic contributions. We present a probabilistic derivation of the algorithm and demonstrate its performance using a simulation, yielding a high quality reconstruction of the Galactic Faraday rotation foreground, a precise estimate of the typical extragalactic contribution, and a well-defined probabilistic description of the extragalactic contribution for each data point. We then apply this reconstruction technique to a catalog of Faraday rotation observations for extragalactic sources. The analysis is done for several different scenarios, for which we consider the error bars of different subsets of the data to accurately describe the observational uncertainties. By comparing the results, we argue that a split that singles out only data near the Galactic poles is the most robust approach. We find that the dispersion of extragalactic contributions to observed Faraday depths is most likely lower than $\SI{7}{\radian/m^2}$, in agreement with earlier results, and that the extragalactic contribution to an individual data point is poorly constrained by the data in most cases.}

\keywords{Magnetic fields -- Methods: data analysis -- ISM: magnetic fields}

\maketitle

\section{Introduction}

Polarized radiation from an astronomical source undergoes Faraday rotation as it travels through the magneto-ionic medium between the source and observer. For extragalactic sources, there are contributions from the Galactic interstellar medium, from any intergalactic magnetic fields, from intervening galaxies on the line of sight, as well as from magnetic fields in the source itself. In this work we attempt to estimate the contribution to the observed Faraday rotation of such sources that is due to magnetic fields outside of the Milky Way. This extragalactic contribution holds the potential for extracting information about cosmic magnetic fields on large scales, e.g., in galaxy clusters, galaxy filaments, or cosmic voids \citep{kolatt-1998, blasi-1999, xu-2006, hammond-2012, bernet-2012, neronov-2013, joshi-2013}.

For a hypothetical source of linear polarization that is point-like in all dimensions and situated at a physical distance $r$ from the observer, the change in polarization angle is given by
\begin{equation}
	\Delta_\chi = \phi \, \lambda^2,
\end{equation}
where $\lambda$ is the wavelength of the radiation and
\begin{equation}
	\label{eq:def_phi}
	\phi = \frac{\mathrm{e}^3}{2\pi \, m_\mathrm{e}^2 \, c^4} \int_r^0 \mathrm{d}r' \, \frac{1}{\left(1 + z(r')\right)^2} \, n_\mathrm{e}{(r')} \, B_r{(r')}
\end{equation}
is the Faraday depth of the source \citep[e.g.,][]{burn-1966}. In the last equation, $n_\mathrm{e}$ is the density of thermal electrons, $B_r$ is the magnetic field vector projected onto the line of sight, $z$ is the cosmological redshift, and the prefactor is a function of the electron charge $\mathrm{e}$, the electron mass $m_\mathrm{e}$, and the speed of light $c$.

The line-of-sight integral in Eq.~\eqref{eq:def_phi} can be split into an integral over the portion of the line of sight that lies within the Milky Way and the portion that is outside the Milky Way, i.e., the Galactic and extragalactic contributions.

For most extragalactic sources, the net rotation is dominated by the effect of the interstellar medium of the Milky Way \citep{leahy-1987, schnitzeler-2010}. This Galactic contribution has recently been estimated from a collection of observations of Faraday rotation of extragalactic sources \citep{oppermann-2012}. One way of estimating the extragalactic contributions is to subtract the estimate of the Galactic contribution from the observed values. However, we will argue that this is not a good estimate due to the presence of uncertainties both in the observations and in the foreground estimate.

Extracting the sub-dominant extragalactic part from the data is more difficult than extracting the Galactic part for two reasons. The first obvious reason is that, as it is smaller, it is more heavily obscured by observational noise in the data. In fact, for many of the data points that we use the expected extragalactic contribution is comparable to the observational uncertainty. The second reason is that the Galactic foreground contribution is spatially smooth to some extent, which enables the usage of correlation information and thus the transfer of information from many data points to each location on the sky. The extragalactic contributions, on the other hand, are expected to be mostly uncorrelated from source to source, given the typical source separation. Therefore, information on the extragalactic contribution to a data point is only contained in the other data points indirectly via the common Galactic foreground. Furthermore, the measurement errors are uncorrelated from source to source as well leading to a  statistical degeneracy with the extragalactic contributions. This means that any split between extragalactic contributions and observational noise in the estimate can only be made according to the expected variances of these two components.

We will additionally argue that the statistical characterization of the observational uncertainties given by the error bars in the data catalogs may be incomplete in some cases. Therefore, there is an additional degree of freedom in the expected noise variance that further exacerbates the degeneracy between observational noise and extragalactic contributions.

The dispersion of the extragalactic contributions has previously been estimated by \citet{schnitzeler-2010}, who studied the spread of the distribution of observed Faraday depths of extragalactic sources from the catalog of \citet{taylor-2009}. \citet{schnitzeler-2010} observed that this spread changes as a function of Galactic latitude even after the subtraction of a coarse foreground model. He then extracted an upper bound on the spread of the extragalactic contributions as the latitude independent part of this function. Here we will regard this spread as unknown and derive a complementary estimate.

For this, we will separate the data into a subset for which the uncertainty information is complete and reliable and a subset for which this is not necessarily the case. The reconstruction of the dispersion of the extragalactic contributions will then be driven mainly by the former subset of the data. We will argue that the best choice for the first subset consists of data that have not only well-described observational uncertainties, but also a small Galactic foreground contribution, i.e., we will prefer data from the Galactic polar regions. The Galactic contribution will be separated off by considering its different spatial correlation structure.

In the following Sect.~\ref{sec:terminology}, we give precise definitions for the terminology that we use in the later discussion. Terms like noise and estimate are introduced and we discuss the relevant probability densities. In Sect.~\ref{sec:algorithm}, we sketch the derivation of the reconstruction algorithm and discuss the degeneracy between the extragalactic and noise contributions. We test the algorithm on a simulation. In discussing the resulting estimates for the extragalactic contributions, we point out the important difference between an estimate and the reality, as well as the importance of considering estimates together with their uncertainties. In Sect.~\ref{sec:realworld}, we apply the algorithm to observational data and present the results. We perform different case studies to gauge the robustness of these results. Finally, we give a brief summary in Sect.~\ref{sec:conclusions}.

A reader interested mainly in the results may skip the derivation and test of the method in Sect.~\ref{sec:algorithm} and go straight from Sect.~\ref{sec:terminology} to Sect.~\ref{sec:polarcapsresults}. In that section, we discuss the results for the split of the data described as the second split in Sect.~\ref{sec:possiblesplits}, which we argue gives the most reliable results. A discussion of the use of these results is given in Appendix~\ref{app:website}.

\section{Data model and terminology}
\label{sec:terminology}

The observed Faraday depth of the $i$-th source in a catalog, or data point $d_i$, is comprised of a Galactic contribution $\phi_{\mathrm{g},i}$, an extragalactic contribution $\phi_{\mathrm{e},i}$, and observational noise $n_i$,
\begin{equation}
    \label{eq:datamodel-general}
    d_i = \phi_{\mathrm{g},i} + \phi_{\mathrm{e},i} + n_i.
\end{equation}
This equation holds for each data point, or equivalently, for all data points at once when one summarizes as vectors $d$ and $n$ the observational estimates and their uncertainties for each source and writes $\phi_{\mathrm{g}/\mathrm{e}}$ as vectors containing the Galactic and extragalactic contributions to the Faraday rotation along all these lines of sight, respectively.

Before discussing possible ways of estimating individual constituents, we introduce the terminology that we will use in our discussion and elaborate on the involved probability densities.

\subsection{Terminology}

We denote the numbers giving the \emph{true} Galactic and extragalactic contributions as $\phi_{\mathrm{g}/\mathrm{e}}$. The \emph{definition} of the noise term, $n$, is then simply the difference between the measured Faraday depth, $d$, and the sum of these two numbers, according to Eq.~\eqref{eq:datamodel-general}. Of these four numbers, the measurement $d$ itself is the only one that is known exactly. In fact, without further input, the three constituents $\phi_\mathrm{g}$, $\phi_\mathrm{e}$, and $n$ are completely degenerate and therefore completely unknown.

Using additional input, which we will discuss in the following, it may be possible to construct reasonable ways to estimate the three numbers adding up to the measured number. We will use hatted variables to denote such estimated quantities, e.g., $\hat{\phi}_\mathrm{g}$ will be an estimate of the Galactic contribution. Distinguishing between the true numbers realized in nature and our estimates of these numbers is crucial.

Noise can arise due to instrumental effects, features of the data processing, or the presence of any other physical effect that is not part of the sum $\left(\phi_\mathrm{g} + \phi_\mathrm{e}\right)$, such as ionospheric Faraday rotation or emission from several Faraday depths on the same line of sight\footnote{\citet{macquart-2012} analysed a subset of the sample of \citet{feain-2009} and found evidence for this complication in 3\,\% of the cases.}. Since the exact contributions of these effects are unknown, we describe the noise via a probability density function (PDF), $\mathcal{P}(n_i|\Theta)$, which gives the probability for the noise contribution to the $i$-th data point to take on a value within the infinitesimal interval between $n_i$ and $n_i + \mathrm{d}n_i$, given a set of assumptions that we make about the processes generating the noise or an effective parameterization of these, denoted as $\Theta$. This PDF immediately provides us with the \emph{likelihood}, i.e., the probability to measure a certain value, assuming certain values for the Galactic and extragalactic contributions (plus the set of assumptions $\Theta$),
\begin{equation}
    \mathcal{P}(d_i|\phi_{\mathrm{g},i},\phi_{\mathrm{e},i},\Theta) = \mathcal{P}(n_i = d_i - \phi_{\mathrm{g},i} - \phi_{\mathrm{e},i}|\Theta).
\end{equation}
This equality holds since, for fixed Galactic and extragalactic contributions, the measurement is completely determined by its noise contribution.

Modeling all effects that can contribute to the noise is often not practical. Therefore, one usually finds a few effective parameters that approximately describe the PDF for the noise contribution or the likelihood. The most common choice, arising e.g., from the central limit theorem or a maximum entropy argument \citep[see, e.g.,][]{jaynes-2003}, is a Gaussian PDF with zero mean, i.e.,
\begin{align}
    \mathcal{P}(d_i|\phi_{\mathrm{g},i},\phi_{\mathrm{e},i},\sigma_i) &= \left(2\pi \sigma_i^2\right)^{-1/2} \exp\left(-\frac{\left(d_i - \phi_{\mathrm{g},i} - \phi_{\mathrm{e},i}\right)^2}{2 \sigma_i^2}\right).
\end{align}
Here, we have parameterized the likelihood completely with its standard deviation $\sigma_i$. We will refer to this parameter as an error bar, since it gives the width of the PDF for the noise (or error) contribution to the measured value. In this formula we have assumed that while we have some idea about the expected magnitude of the noise contribution, given by the standard deviation $\sigma_i$, the specific value of $n_i$ is unknown. We are not considering systematic errors that would lead to an offset in the observational values and thus a non-zero mean of the PDF for the noise.

Precise knowledge of the likelihood function is crucial for the inference algorithms that we will discuss later. However, as we discuss in Appendix \ref{app:npi}, the likelihood is not always well described by the single parameter $\sigma_i$. To accommodate the possibility of having such imperfect error information (among other things), \citet{oppermann-2012} allowed their algorithm to widen the likelihood functions of individual data points considerably. This was warranted because even when the uncertainty information for the vast majority of the data points is reliable, a few poorly described outliers can greatly influence the reconstruction. This was shown by \citet{oppermann-2011} for the method employed by \citet{oppermann-2012} and is still true for the algorithm we are deriving in Sect.~\ref{sec:full}, although to a lesser extent because in our current paper we make use of a spectral smoothness prior that was not used in previous papers.

\subsection{Probability densities}

A generalization of the one-dimensional Gaussian model discussed so far is a multi-dimensional Gaussian model with correlations, described by
\begin{align}
    \mathcal{P}(n|N) &= \mathcal{G}(n,N)\nonumber\\
    &= \frac{1}{\left|2\pi N\right|^{1/2}} \exp\left(-\frac{1}{2} n^\dagger N^{-1} n\right).
\end{align}
Here, the dagger denotes a transposed quantity and we introduced the notation $\mathcal{G}{(x,X)}$ for a Gaussian distribution for a variable $x$ with zero mean and covariance $X$. The covariance matrix $N$ contains the variances of the noise contributions to the individual measurements on its diagonal, $N_{ii} = \sigma_i^2$, and their correlations as off-diagonal entries. For uncorrelated noise contributions, i.e., $N_{ij} = 0$ for $i\neq j$, the one-dimensional Gaussian for each measurement is recovered.

So far, we have only discussed the likelihood. However, since we are trying to infer the Galactic and extragalactic contributions from the measurements and the likelihood is the PDF for the former quantity under the assumption of fixed values for the latter, it is clearly not the PDF that we are interested in. In order to turn the argument around, we make use of Bayes' theorem to construct the \emph{posterior} PDF,
\begin{equation}
    \mathcal{P}(\phi_\mathrm{g},\phi_\mathrm{e}|d,N) \propto \mathcal{P}(d|\phi_\mathrm{g},\phi_\mathrm{e},N) \, \mathcal{P}(\phi_\mathrm{g},\phi_\mathrm{e}),
\end{equation}
which is the PDF for the Galactic and extragalactic contributions to all measurements, given the measured data and the assumptions about the noise covariance. The last PDF on the right hand side is the \emph{prior} PDF for the Galactic and extragalactic contributions, i.e., a summary of knowledge we have about these constituents before taking into account the measurement data. For example, a prior could encode information about the expected variability or spatial smoothness of the Galactic and extragalactic contributions and thus serve to break the degeneracy between the two.

The posterior PDF encodes all the knowledge that is available about the quantities of interest, both from the measurement data and from prior assumptions. Therefore, the posterior PDF should be the main result of an analysis. However, it may be practical to summarize the information. To this end, we will approximate the posterior in our analysis as a Gaussian. This Gaussian is described by the \emph{posterior mean}
\begin{equation}
    \hat{\phi}_{\mathrm{g}/\mathrm{e}} = \left<\phi_{\mathrm{g}/\mathrm{e}}\right>_{(\phi_\mathrm{g/e}|d)},
\end{equation}
a weighted average of all possible contributions, and the posterior covariance,
\begin{equation}
    D_\mathrm{g/e} = \left<\left(\phi_\mathrm{g/e} - \hat{\phi}_\mathrm{g/e}\right)\left(\phi_\mathrm{g/e} - \hat{\phi}_\mathrm{g/e}\right)^\dagger \right>_{(\phi_\mathrm{g/e}|d)},
\end{equation}
which encodes information on the uncertainty of the estimate. We use the angle bracket notation to denote integrals over all possible configurations of the quantity given in the index,
\begin{equation}
    \left< f(x) \right>_{(x|X)} = \int\mathcal{D}x \, f(x) \, \mathcal{P}(x|X).
\end{equation}

As a prior, we will use a multi-dimensional Gaussian distribution with no linear correlation between any pair of the three constituents. This does not exclude a correlation between, e.g., the noise \emph{variance} and the Galactic contribution.

\section{Reconstruction algorithm}
\label{sec:algorithm}

As we already pointed out, an exact and definitive separation of $d$ into its three constituents is not possible. However, under reasonable assumptions, we can formulate estimates for these quantities and draw conclusions about the statistics of the extragalactic contributions. In this section, we will develop an algorithm for calculating such an estimate step by step. We begin by discussing the problem for known covariance matrices, then describe the assumptions that we make about the symmetries of these covariance matrices, and finally outline the derivation of the complete algorithm and test it on a simulation.

\subsection{Wiener filter}
\label{sec:Wienerfilter}

To estimate the Galactic contribution to Faraday rotation, \citet{oppermann-2012} used an algorithm based on the Wiener filter, which calculates the linear estimate, $\hat{\phi}_\mathrm{g} = Fd$, that minimizes the expectation value of the square-norm of the residual $r = \phi_\mathrm{g} - \hat{\phi}_\mathrm{g}$. Here, the expectation value is calculated over the joint PDF of all constituents of the data, so that $F$ minimizes the expression
\begin{equation}
    \label{eq:minimization}
    \left<\left(\phi_\mathrm{g} - Fd\right)^\dagger \left(\phi_\mathrm{g} - Fd\right)\right>_{(\phi_\mathrm{g},\phi_\mathrm{e},n)}.
\end{equation}
The Wiener filter reconstruction $Fd$ indeed yields the optimal estimate in this square-norm sense in cases in which all priors are Gaussian, since it equals the posterior mean of $\phi_\mathrm{g}$ in this case \citep[see, e.g.,][]{ensslin-2010}. In all other cases, the Wiener filter is still the optimal \emph{linear} filter.

Solving for the filter matrix $F$ yields \citep[see, e.g.,][]{zaroubi-1995}
\begin{equation}
    \label{eq:optlinfilt}
    F = \left<\phi_\mathrm{g} d^\dagger\right>_{(\phi_\mathrm{g},\phi_\mathrm{e},n)} \left<dd^\dagger\right>_{(\phi_\mathrm{g},\phi_\mathrm{e},n)}^{-1}.
\end{equation}
Assuming all three constituents of the data to be mutually linearly uncorrelated, the expectation values in the last equation simplify to
\begin{equation}
    \left<\phi_\mathrm{g} d^\dagger\right>_{(\phi_\mathrm{g},\phi_\mathrm{e},n)} = \left<\phi_\mathrm{g} \phi_\mathrm{g}^\dagger\right>_{(\phi_\mathrm{g})} \equiv G
\end{equation}
and
\begin{equation}
    \left<dd^\dagger\right>_{(\phi_\mathrm{g},\phi_\mathrm{e},n)} = \left<\phi_\mathrm{g} \phi_\mathrm{g}^\dagger\right>_{(\phi_\mathrm{g})} + \left<\phi_\mathrm{e} \phi_\mathrm{e}^\dagger\right>_{(\phi_\mathrm{e})} + \left<n n^\dagger\right>_{(n)} \equiv G + E + N,
\end{equation}
where we have introduced the covariance matrices $G$, $E$, and $N$ of the Galactic, extragalactic, and noise contributions, respectively. With these abbreviations, the Wiener filter estimates for the three constituents become
\begin{equation}
    \label{eq:galest}
    \hat{\phi}_\mathrm{g} = G \left( G + E + N \right)^{-1} d,
\end{equation}
\begin{equation}
    \label{eq:exest}
    \hat{\phi}_\mathrm{e} = E \left( G + E + N \right)^{-1} d = E \left(E + N\right)^{-1} \left(d - \hat{\phi}_\mathrm{g}\right),
\end{equation}
\begin{equation}
    \label{eq:noiseest}
    \hat{n} = N \left( G + E + N \right)^{-1} d = N \left(E + N\right)^{-1} \left(d - \hat{\phi}_\mathrm{g}\right).
\end{equation}

It should be noted that Eq.~\eqref{eq:exest} is not the same as taking the difference between the data and the Wiener filter estimate for the Galactic contribution. This difference would rather be an optimal estimate of the sum of the extragalactic and noise contributions and we get the desired estimate by weighting it with the ratio of the expected extragalactic variance to the expected non-galactic variance as written out on the right hand side of Eq.~\eqref{eq:exest}.

\subsubsection{Illustration}
\label{sec:illustration}

\begin{figure}
    \includegraphics[width=\columnwidth]{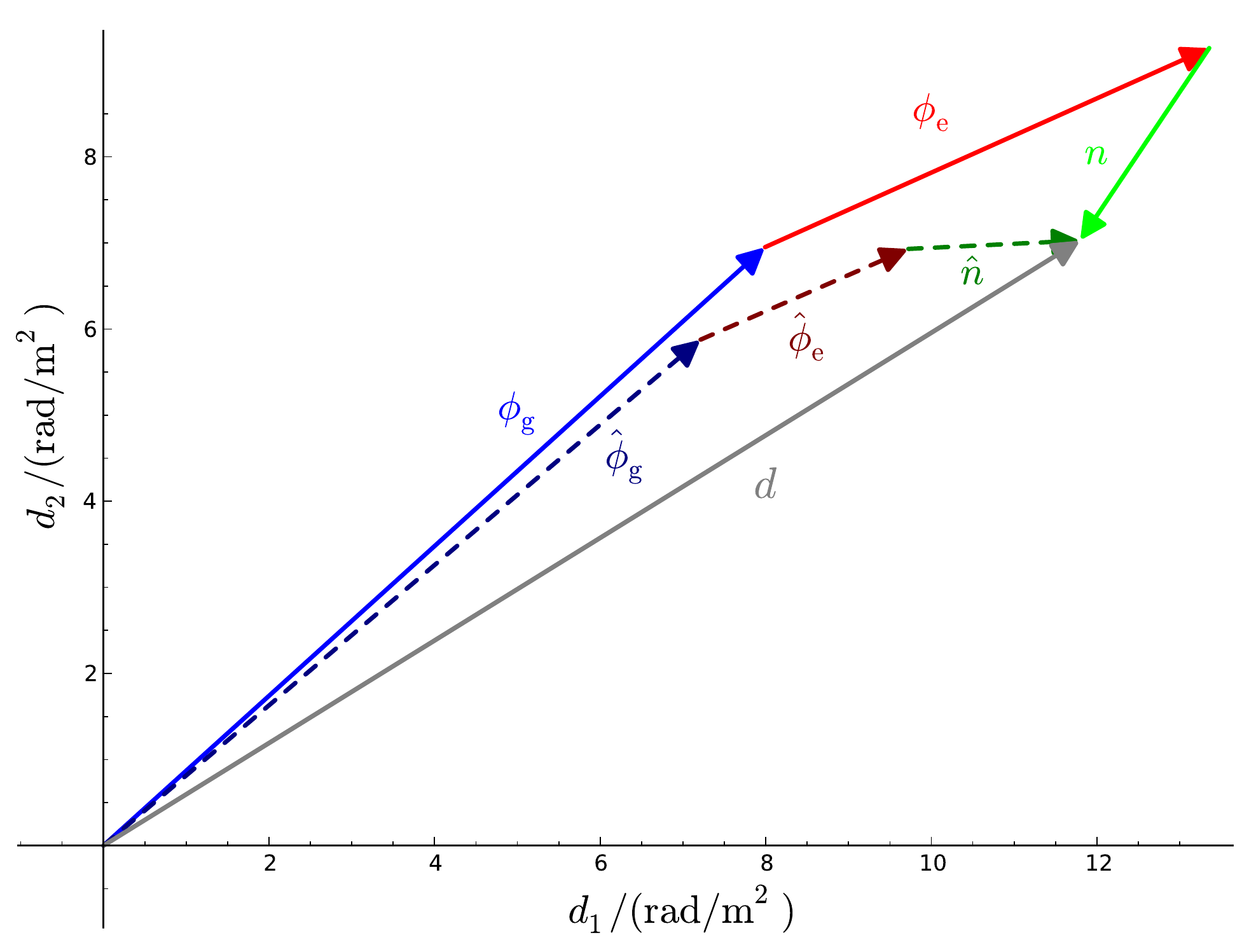}
    \caption{\label{fig:vectors}Visualization of the data space in the case of only two Faraday rotation measurements. Each coordinate denotes the possible values for the two measurements $d_{1}$ and $d_{2}$ at two sky positions. The data vector, $d=(d_{1,}\, d_{2})^{\dagger}$ is the sum of a vector of Galactic Faraday depth, $\phi_\mathrm{g}=(\phi_{\mathrm{g},1}, \phi_{\mathrm{g},2})^{\dagger}$, of a vector of extragalactic Faraday depth, $\phi_\mathrm{e}=(\phi_{\mathrm{e},1}, \phi_{\mathrm{e},2})^{\dagger}$, and of a measurement noise vector, $n=(n_{1}, n_{2})^{\dagger}$. The reconstructions of these three components, $\hat{\phi}_\mathrm{g}$, $\hat{\phi}_\mathrm{e}$, and $\hat{n}$, are shown as dashed arrows. Their sum is also equal to the data, however they differ from the correct components due to the impossibility to uniquely separate one data vector into three statistically independent components.}
\end{figure}

To illustrate how the Wiener filter gives a non-trivial estimate for all three constituents, we investigate the case of only two Faraday rotation measurements at two different locations on the sky. In this case the data vector is two-dimensional, $d=(d_{1,}\, d_{2})^{\dagger}$, and can easily be plotted as a vector, see Fig.~\ref{fig:vectors}. We assume the data according to Eq.~\eqref{eq:datamodel-general} to be the result of three independent stochastic processes, namely Galactic and extragalactic Faraday rotation at these locations plus measurement noise. We model these to be Gaussian processes with covariances
\begin{align}
    G &=\begin{pmatrix}(10)^{2} & 60\\
    60 & (10)^{2}
    \end{pmatrix} \SI{}{\radian^2/m^4},\\
    E &=\begin{pmatrix}(6.6)^{2} & 0\\
    0 & (6.6)^{2}
    \end{pmatrix} \SI{}{\radian^2/m^4},\mbox{ and}\\
    N &=\begin{pmatrix}(6.0)^{2} & 0\\
    0 & (2.0)^{2}
    \end{pmatrix} \SI{}{\radian^2/m^4}.
\end{align}

The noise of the two measurements is independent and is assumed in our example to have a standard deviation of $\SI{6}{\radian/m^2}$ and $\SI{2}{\radian/m^2}$ for the two different observations, respectively. The extragalactic contributions have the same variance everywhere, with a standard deviation of $\SI{6.6}{\radian/m^2}$ \citep[based on][]{schnitzeler-2010}. The Galactic components on the two measurement locations are assumed to be correlated, leading to a non-diagonal correlation matrix $G$. In Fig.~\ref{fig:vectors} we show the component vectors drawn to generate the data as well as their reconstruction from the data according to Eqs.~\eqref{eq:galest}, \eqref{eq:exest}, and \eqref{eq:noiseest} under the assumption of known covariance matrices. These estimates add up to the data vector, $\hat{\phi}_\mathrm{g} + \hat{\phi}_\mathrm{e} + \hat{n} = d = \phi_\mathrm{g} + \phi_\mathrm{e} + n$, without being identical to the original signals. Due to the different structure of the covariance matrices it is possible to construct an optimal and non-trivial partition of the data vector in non-parallel components.

This example shows that the reconstruction is able to capture some aspects of the components, but not all. In particular it illustrates that the reconstructed quantities have less variance than the original ones. The estimates $\hat{\phi}_\mathrm{e}$ and $\hat{n}$ simply split the difference $d - \hat{\phi}_\mathrm{g}$ according to the ratio of the variances, $(6.6)^2/(6.0)^2$ and $(6.6)^2/(2.0)^2$ for each of the two data points, respectively. This is due to both $E$ and $N$ being diagonal. Furthermore, Fig.~\ref{fig:vectors} also illustrates that using $d - \hat{\phi}_\mathrm{g}$ as an estimator for $\phi_\mathrm{e}$ is suboptimal, since it contains some of the noise, as we have $d - \hat{\phi}_\mathrm{g} = \hat{\phi}_\mathrm{e} + \hat{n}$. 

The amount of variance missing from the estimates is known statistically and can be characterized by the covariance matrices $D_\mathrm{g} = \left(G^{-1}+(N+E)^{-1}\right)^{-1}$, $D_\mathrm{e} = \left(E^{-1}+(N+G)^{-1}\right)^{-1}$, and $D_{n} = \left(N^{-1}+(G+E)^{-1}\right)^{-1}$ for the Galactic, extragalactic, and noise components, respectively.

\subsection{Filtering with unknown covariance matrices}
\label{sec:full}

The main problem with our considerations so far is that the covariance matrices are not necessarily known with sufficient precision. In this section we extend the analysis of Sect.~\ref{sec:Wienerfilter} to a reconstruction with unknown covariance matrices. In the first subsection, we lay out the necessary assumptions about the symmetries of the covariance matrices and the priors for the remaining parameters. We then present the derivation of the algorithm from probabilistic considerations, demonstrate its performance in a simulated scenario, and finally show the results for the analysis of real-world data in Sect.~\ref{sec:realworld}.

\subsubsection{Assumptions and covariance matrices}
\label{sec:covariances}

In the previous section we have argued for the Wiener filter as the optimal linear estimate and have found that the covariance matrices $G$, $E$, and $N$ play an important role. Here, we will go further and explicitly assume Gaussian priors for all three constituents, i.e., we make use of only the covariance matrices to describe the prior statistics \citep[see, e.g.,][for a discourse on the appropriateness of Gaussian priors in such a case]{jaynes-2003}.

\paragraph{Galactic covariance}

For the prior covariance of the Galactic contribution, we follow the argument of \citet{oppermann-2012}. We model the Galactic Faraday depth as a dimensionless isotropic Gaussian random field, $s(l,b)$, multiplied with a latitude-dependent profile function, $p(b)$, i.e.,
\begin{align}
	\label{eq:intro_p}
    \phi_\mathrm{g}(l_i,b_i) &= p(b_i) \, s(l_i,b_i)\nonumber\\
    &= \int_{\mathcal{S}^2} \mathrm{d}\Omega ~ R_i(l,b) \, s(l,b),
\end{align}
where $l_i$ and $b_i$ are the Galactic longitude and latitude of the $i$-th data point and we have introduced a response operator $R$ that includes a projection of the all-sky field $s$ onto the locations of the observed sources and a multiplication with the profile function $p$. With this response operator we can write the covariance matrix of the Galactic contribution, regarded as a quantity defined only in the locations of observed sources, as
\begin{equation}
    G = RSR^\dagger.
\end{equation}
Inferring the matrix $G$ from the data is part of the inference problem.

Modeling the field $s$ as an isotropic field means that we can write its covariance matrix as diagonal in the basis of spherical harmonics components according to
\begin{equation}
    S_{(\ell m),(\ell' m')} = \left<s_{\ell m} s_{\ell' m'}^* \right>_{(s|S)} = \delta_{\ell \ell'} \, \delta_{mm'} \, C_\ell.
\end{equation}
Here, $\delta$ is the Kronecker delta symbol, the asterisk denotes complex conjugation, and $C_\ell$ denotes the angular power spectrum on scale $\Delta_\theta \approx \SI{180}{\degree}/\ell$.

We constrain the angular power spectrum with a prior consisting of two parts. For the first part we still follow \citet{oppermann-2012} and use an independent inverse-Gamma prior for each component,
\begin{equation}
	\label{eq:IG-prior}
    \mathcal{P}_\mathrm{IG}(C_\ell) \propto C_\ell^{-\alpha_\ell} \exp{\left(-\frac{q_\ell}{C_\ell}\right)},
\end{equation}
where $q_\ell$ and $\alpha_\ell$ are parameters that vary the constraining power of the prior. The second part of our power spectrum prior is a term that couples different scales and enforces spectral smoothness. This term is given by
\begin{align}
    \label{eq:smoothnessprior}
    \mathcal{P}_\mathrm{sm}{(C)} &\propto
	\exp{\left(-\frac{1}{2\sigma_\mathrm{sm}^2}
      \int\mathrm{d}{(\log\ell)} ~ \left(\frac{\partial^2 \log
        C_\ell}{\partial{(\log\ell)^2}}\right)^2\right)} \nonumber\\
    &= 
    \exp{\left(-\frac{1}{2}(\log C)^\dagger T (\log C) \right)},
\end{align}
where the second derivative and integral are to be read as short-hands
for finite-difference expressions and $T$ is a matrix that performs
the same operations. This prior favors angular power spectra that are close to power laws. Its strength is regulated by the parameter $\sigma_\mathrm{sm}$. In total, the prior for the angular power spectrum is the product of the two terms,
\begin{equation}
	\label{eq:prior-C}
    \mathcal{P}(C) \propto \mathcal{P}_\mathrm{sm}{(C)} \, \prod_\ell \mathcal{P}_\mathrm{IG}{(C_\ell)}.
\end{equation}
We refer the reader to \citet{oppermann-2013} for a detailed discussion of this spectral prior.

\paragraph{Extragalactic and noise covariances}

The extragalactic and noise contributions can both be regarded as quantities that are defined only for the source positions. For the noise this choice is obvious since there is no noise if there is no measurement. For the extragalactic contribution the data have no constraining power at any other locations with the prior assumptions that we are about to make and we therefore only consider the extragalactic contributions at the source positions for the sake of simplicity.

For the extragalactic and noise contributions, we choose covariance matrices $E$ and $N$ that are diagonal, i.e., the extragalactic and noise contributions to the individual measurements are regarded as uncorrelated. In principle, a coherent magnetic field in the intergalactic medium within the Local Group or on cosmological scales could cause a correlated extragalactic Faraday rotation structure \citep[e.g.,][]{kolatt-1998, blasi-1999, xu-2006}. However, \citet{akahori-2011} have shown that this contribution is expected to be about 3 to 4 orders of magnitude smaller than the contributions intrinsic to the sources and due to large scale structure filaments.

Since these two matrices are diagonal in the same basis, their entries are degenerate if both matrices are unknown. To break this degeneracy, we will assume that some of the entries of the noise covariance matrix are in fact known with certainty a priori, which will enable us to extract the variance of the extragalactic contributions. We will refer to these data as \emph{Very Informative Points} (VIP). The remaining data, for which we cannot be certain that the quoted error bar describes the likelihood function accurately enough, will be referred to as \emph{Somewhat Informative Points} (SIP). We then model the covariances of the contributions to VIP data individually as
\begin{equation}
    \label{eq:noisegooddata}
    N_{jj} = \sigma_j^2,
\end{equation}
\begin{equation}
    \label{eq:exgooddata}
    E_{jj} = \eta_\mathrm{e} \sigma_\mathrm{e}^2,
\end{equation}
and the sum of the covariances for the SIP data as
\begin{equation}
    \label{eq:exandnoisebaddata}
    \left(E + N\right)_{ii} = \eta_i \left(\sigma_\mathrm{e}^2 + \sigma_i^2\right).
\end{equation}
Here and in the following, $\sigma_\mathrm{e}$ is an initial guess for the dispersion of the extragalactic contributions, which is the same for all data points, and $\sigma_i$ denotes the observational error bar for the $i$-th data point. The factor $\eta_i$ is a correction factor that is introduced to account for possible insufficiencies in either or both of the two estimates, i.e., we do not attempt to differentiate between insufficient error information and an under-estimated variance of the extragalactic contribution. For the VIP data we do not allow for corrections of the error bars but use them to correct our initial assumption about the typical variance of the extragalactic contributions via the factor $\eta_\mathrm{e}$. The error variance correction factors in Eq.~\eqref{eq:exandnoisebaddata} are individually determined for each data point, whereas the factor $\eta_\mathrm{e}$ in Eq.~\eqref{eq:exgooddata} is the same for each data point in the VIP category, since we assume that the statistics of the extragalactic contributions are homogeneous over the whole sky.

In the following, we denote as $d^\mathrm{(SIP)}$, $\phi_\mathrm{e}^\mathrm{(SIP)}$, $n^\mathrm{(SIP)}$ the vectors containing the observed values, extragalactic contributions, and noise contributions of the data points in the SIP category, described by Eq.~\eqref{eq:exandnoisebaddata}, and as $d^\mathrm{(VIP)}$, $\phi_\mathrm{e}^\mathrm{(VIP)}$, $n^\mathrm{(VIP)}$ the same for the VIP category of data, described by Eqs.~\eqref{eq:noisegooddata} and \eqref{eq:exgooddata}. In Sect.~\ref{sec:realworld} we will present several possible choices for splitting the data into the VIP and SIP categories.

For the free parameters $\eta_i$ and $\eta_\mathrm{e}$, we choose again inverse-Gamma priors,
\begin{equation}
	\label{eq:prior-eta}
    \mathcal{P}_\mathrm{IG}{(\eta_{i/\mathrm{e}})} \propto \eta_{i/\mathrm{e}}^{-\beta_{i/\mathrm{e}}} \exp{\left(-\frac{r_{i/\mathrm{e}}}{\eta_{i/\mathrm{e}}}\right)}.
\end{equation}
We will discuss our choice of parameters $r_{i/\mathrm{e}}$ and
$\beta_{i/\mathrm{e}}$ in the following section. Naturally, we do not enforce smoothness of the $\eta$-values in any sense.

The choice of inverse-gamma priors and their parameters, as well as the specific form and strength of the spectral smoothness prior, are arbitrary to some degree. The simulation we study in the next section, however, will show that the choice we make here yields reasonable results.

\subsubsection{Reconstruction strategy}
\label{sec:strategy}

Here we make use of the combined methodology of
\citet{oppermann-2011} and \citet{oppermann-2013}. In the first of
these papers a method for the reconstruction of a Gaussian signal
field, its power spectrum, and the $\eta$-factors was presented, the
so-called \emph{extended critical filter}. In
the second paper, the inclusion of a spectral smoothness prior
according to Eq.~\eqref{eq:smoothnessprior} was discussed, albeit
without allowing for the $\eta$-factors. Here we combine these two methods and add the split into two qualitatively different data categories using the ansatz of the empirical Bayes method
\citep[e.g.,][]{robbins-1955}. We approximate the posterior mean for
the dimensionless signal field and the extragalactic contributions as
\begin{align}
    m &= \int\mathcal{D}s \int\mathcal{D}\tilde{C} \int \mathcal{D}\tilde{\eta} ~ s \, \mathcal{P}{(s|d,\tilde{C},\tilde{\eta})}
    \, \mathcal{P}{(\tilde{C},\tilde{\eta}|d)}\nonumber\\
    &\approx \int\mathcal{D}s \int\mathcal{D}\tilde{C} \int \mathcal{D}\tilde{\eta} ~ s \,
      \mathcal{P}{(s|d,\tilde{C},\tilde{\eta})} \, \delta{(\tilde{C} -
        \hat{\tilde{C}})} \, \delta{(\tilde{\eta} -
        \hat{\tilde{\eta}})},\\
    \hat{\phi}_\mathrm{e} &= \int\mathcal{D}\phi_\mathrm{e} \int\mathcal{D}\tilde{C} \int\mathcal{D}\tilde{\eta} ~ \phi_\mathrm{e} \,
    \mathcal{P}{(\phi_\mathrm{e}|d,\tilde{C},\tilde{\eta})} \,
    \mathcal{P}{(\tilde{C},\tilde{\eta}|d)}\nonumber\\
    &\approx \int\mathcal{D}\phi_\mathrm{e} \int\mathcal{D}\tilde{C} \int\mathcal{D}\tilde{\eta} ~ \phi_\mathrm{e} \,
      \mathcal{P}{(\phi_\mathrm{e}|d,\tilde{C},\tilde{\eta})}
      \,\delta{(\tilde{C} - \hat{\tilde{C}})} \, \delta{(\tilde{\eta}
        - \hat{\tilde{\eta}})},
\end{align}
where we have chosen to work with the logarithmic quantities $\tilde{C} = \log C$ and $\tilde{\eta} = \log \eta$. We choose as estimators for the auxiliary quantities, $(\hat{\tilde{C}}_\ell)_\ell$, $(\hat{\tilde{\eta}}_i)_i$, and $\hat{\tilde{\eta}}_\mathrm{e}$ the numbers that maximize the PDFs
\begin{equation}
    \mathcal{P}{\left(\left.(\tilde{C}_\ell)_\ell \right| d, (\tilde{\eta}_i)_i = (\hat{\tilde{\eta}}_i)_i, \eta_\mathrm{e} = \hat{\eta}_\mathrm{e}\right)},
\end{equation}
\begin{equation}
    \mathcal{P}{\left(\left.(\tilde{\eta}_i)_i \right| d, (\tilde{C}_\ell)_\ell = (\hat{\tilde{C}}_\ell)_\ell, \eta_\mathrm{e} = \hat{\eta}_\mathrm{e}\right)},
\end{equation}
and
\begin{equation}
	\mathcal{P}{\left(\tilde{\eta}_\mathrm{e} \left| d, (\tilde{C}_\ell)_\ell = (\hat{\tilde{C}}_\ell)_\ell\right.\right)},
\end{equation}
respectively. This leads to estimates $m$ for the signal field and
$\hat{\phi}_\mathrm{e}$ for the extragalactic contributions that depend on the estimates $\hat{\tilde{C}}$ and
$\hat{\tilde{\eta}}$, which are themselves again dependent on each
other and on the estimates $m$ and $\hat{\phi}_\mathrm{e}$. A system of equations arises that needs
to be solved self-consistently. Our estimate for the Galactic contributions to the observed values will then be $\hat{\phi}_\mathrm{g} = Rm$.

Putting in the priors that we described in the previous section, we can calculate the PDFs needed for the estimates, after making a few approximations. The detailed calculations and filter formulas are discussed in Appendix~\ref{app:derivation}.

The choice of estimators made here represents a trade-off between statistical optimality and practical ability to calculate the relevant PDFs. Ideally, we would estimate each quantity by marginalizing over all other unknown quantities and averaging the resulting posterior distribution. However, these marginalizations are in general not possible analytically. Rather than resorting to computationally expensive sampling techniques, we use some unmarginalized PDFs. In the next section, we show a simulated example calculation that demonstrates the quality of the results obtained with our approximations.

We fix the remaining prior parameters according to the following scheme. For the reconstruction of the angular power spectrum of the Galactic contributions, we use the limit $q \rightarrow 0$ and $\alpha \rightarrow 1$. This turns the inverse-gamma prior for each parameter $C_\ell$ into a Jeffreys prior, and makes the prior for $\log C_\ell$ flat. For the strength of the spectral smoothness prior, we choose $\sigma_\mathrm{sm}^2 = 10$, entering the equations via the matrix $T$. This is a rather weak smoothness prior, allowing for a change in slope of $\sqrt{10}$ per e-folding in $\ell$ on a $1\sigma$-level. The reconstruction of the power spectrum will therefore be largely data-driven. For the prior for the noise variance correction factors we choose $\beta_i = 2$, making this inverse-gamma prior more informative than the one used for the angular power spectrum. This is done to account for the expectation that most of the data points in the SIP category will have error bars that describe the likelihood sufficiently well and therefore will not need large $\eta_i$-factors. For the cutoff-parameters $r_i$, we adopt the values
\begin{equation}
	r_i = \frac{3}{2} \max{\left\{ 1, \frac{\sigma_i^2 + \eta_\mathrm{e} \sigma_\mathrm{e}^2}{\sigma_i^2 + \sigma_\mathrm{e}^2} \right\}}.
\end{equation}
The lower threshold of $3/2$ is introduced to make sure that after all approximations made in the derivation, the noise variance correction factors never \emph{decrease} the uncertainty in the measurement, setting $\eta_i = 1$ as a lower limit. This threshold is increased whenever it becomes possible for the pure noise variance of a data point,
\begin{equation}
	N_{ii} = \left(N + E\right)_{ii} - E_{ii} = \eta_i \left( \sigma_i^2 + \sigma_\mathrm{e}^2 \right) - \eta_\mathrm{e} \sigma_\mathrm{e}^2,
\end{equation}
to decrease with respect to the initial value $\sigma_i^2$.

Finally, for the correction factor for the extragalactic variance, we again use a Jeffreys prior, i.e., $\beta_\mathrm{e} \rightarrow 1$ and $r_\mathrm{e} \rightarrow 0$. This prior is broader than needed, since the order of magnitude of the extragalactic variance is already known, e.g., from \citet{schnitzeler-2010}. However, we expect the extragalactic variance to be sufficiently constrained by the data in the VIP category, so that we do not need to constrain it with the prior. As in the case with the spectral smoothness prior, we choose to forgo the use of a stricter prior in favor of a more data-driven analysis.

\subsection{Simulation}
\label{sec:simulation}

To investigate the properties and limitations of the algorithm we developed in the previous section, we will apply it to a simulation of the Faraday sky.

\subsubsection{Simulation setup}

\begin{figure}
    \input{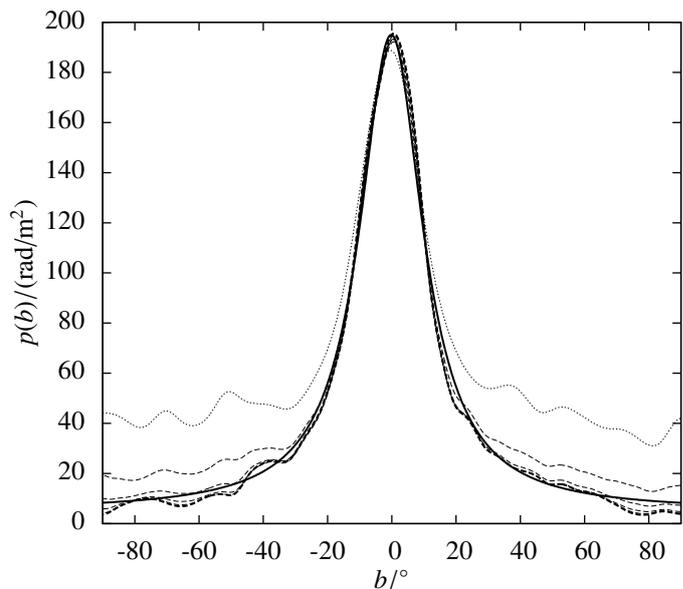}
    \caption{\label{fig:profiles}Galactic latitude profile used in the
      simulation and its reconstruction. The thick solid line is the
      simulated profile, the thin dotted line is the profile initially
      used in the reconstruction and calculated directly from the
      simulated data. Subsequent reconstructions of the profile
      function are shown as thin dashed lines with earlier iterations
      lying higher in the plot. The final reconstruction is depicted
      by the thick dashed line.}
\end{figure}

\begin{figure}
    \input{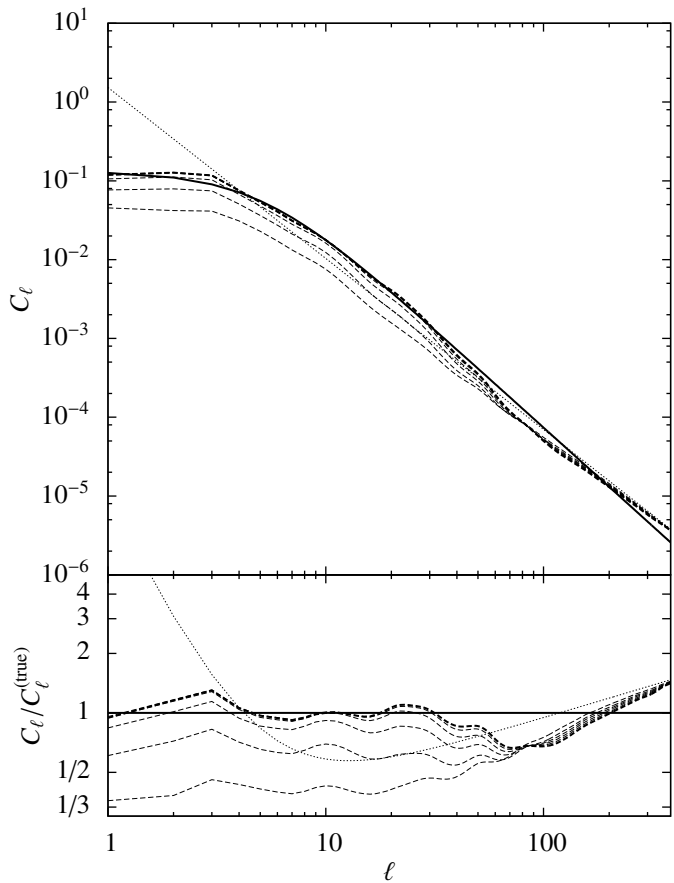}
    \caption{\label{fig:powerspectra}Angular power spectrum of the
      simulated dimensionless version of the Galactic Faraday depth
      and its reconstructions. The thick solid line is the simulated
      power spectrum, the thin dotted line shows the initial guess,
      the thin dashed lines are subsequent reconstructions, and the
      thick dashed line is the final reconstruction. Shown are the
      reconstructions at the end of each iteration with fixed Galactic
    latitude profile. In the bottom panel, we show the ratio of the
    reconstructed power spectra and the one used in the simulation.}
\end{figure}

\begin{figure*}
    \includegraphics[width=\textwidth]{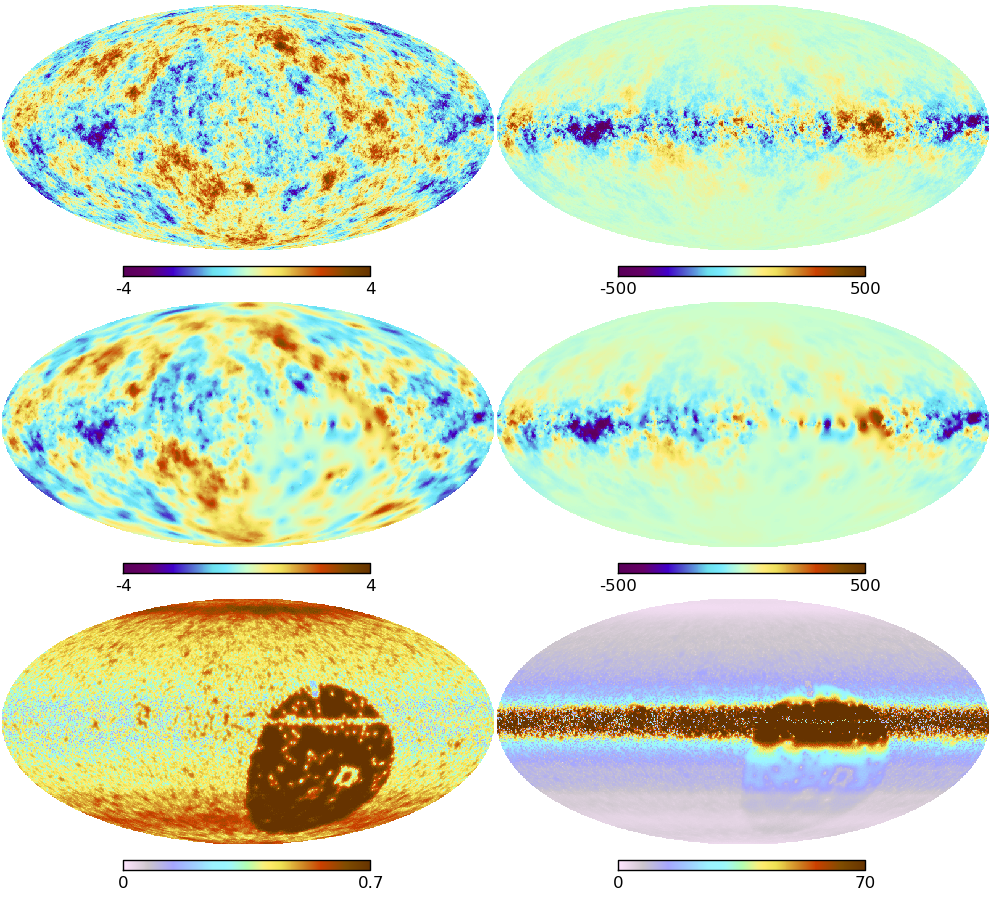}
    \caption{\label{fig:maps}Simulation and reconstruction of the
      Galactic contribution to Faraday rotation. The left column shows
      the dimensionless isotropic Gaussian random field used in our
      statistical model. The right column shows the resulting physical
      Galactic Faraday depth in units of \SI{}{\radian/m^2}. The top row shows the simulation, the middle row its reconstruction, and the bottom row the reconstruction's uncertainty per pixel.}
\end{figure*}

We model the Galactic Faraday depth as a dimensionless, isotropic, correlated Gaussian random field multiplied with a latitude-dependent profile function. The profile function $p(b)$ and angular power spectrum $C_\ell$ that we use are shown in Figs.~\ref{fig:profiles} and \ref{fig:powerspectra}, respectively, with a thick solid line. Our choice of profile function is modeled on the one found by \citet{oppermann-2012} and as a power spectrum we choose a simple broken power-law,
\begin{equation}
    C_\ell \propto \left( 1 + \left(\frac{\ell}{\ell_0}\right)^2 \right)^{-\gamma/2},
\end{equation}
with breaking point $\ell_0 = 5$ and spectral index $\gamma = 2.5$, as an arbitrary model that is not too far removed from what we expect \citep{oppermann-2012}. We choose the normalization of the power spectrum in such a way that the resulting dimensionless field $s(l,b)$ has variance one. This field and the corresponding map of the simulated Galactic Faraday depth are shown in the top two panels of Fig.~\ref{fig:maps}.

The data points we simulate are at the source locations of the catalog used by \citet{oppermann-2012}, which we will again use in Sect.~\ref{sec:realworld}. For the extragalactic contribution to these data points, we draw independent zero-centered Gaussian random numbers with a dispersion of $\sigma_\mathrm{e}^{\mathrm{(true)}} = \SI{10}{\radian/m^2}$. This is higher than the starting value of $\sigma_\mathrm{e} = \SI{6.6}{\radian/m^2}$ that we use in our reconstruction. Checking whether the algorithm is able to pick up this discrepancy is one of the main purposes of this study.

\begin{figure}
    \includegraphics[width=\columnwidth]{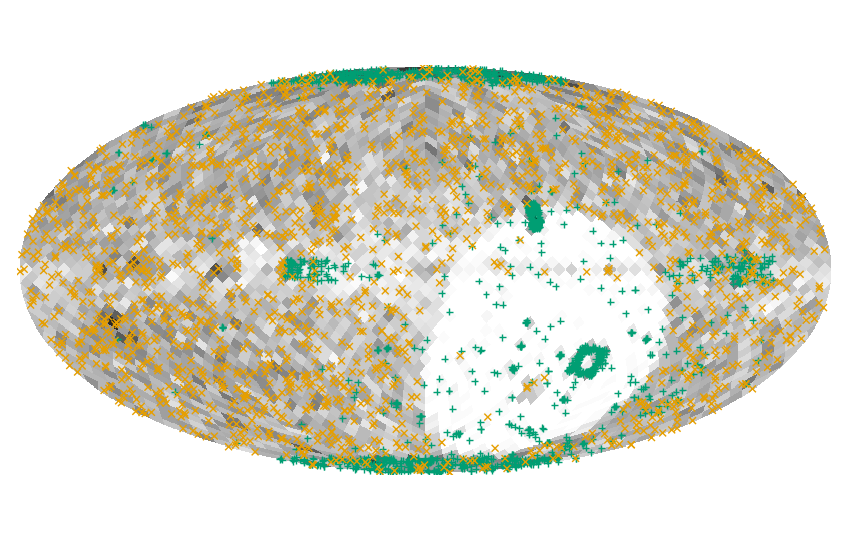}
    \caption{\label{fig:datapoints}Locations of the simulated data
      points on the sky. The orange $\times$-signs denote data points of the
      SIP category for which the noise variance has been
      increased. Data points of the VIP category are shown as green $+$-signs and the grayscale shows the overall density of data points.}
\end{figure}

Finally, we assume uncorrelated Gaussian noise contributions for all data points. The individual noise variances are given by the error bars of the catalog in most cases. However, we do split the data points into two categories, as described earlier, and increase the error variance for a randomly chosen subset of 5\% of the data points in the SIP category by an arbitrary factor of 400 (i.e., we increase the error bar by a factor 20), without informing the reconstruction algorithm about the magnitude and locations of this effect. We split the data into the SIP and VIP categories according to the methodology used in the calculation of the data values, putting all data points derived via a $\lambda^2$-fit into the SIP category and all data points derived using the RM-synthesis technique into the VIP category. Table~1 of \citet{oppermann-2012} includes this information for all data points. The resulting distribution of data points of the two categories, as well as the location of the data points with increased noise variance, is shown in Fig.~\ref{fig:datapoints}.

\begin{figure}
  \input{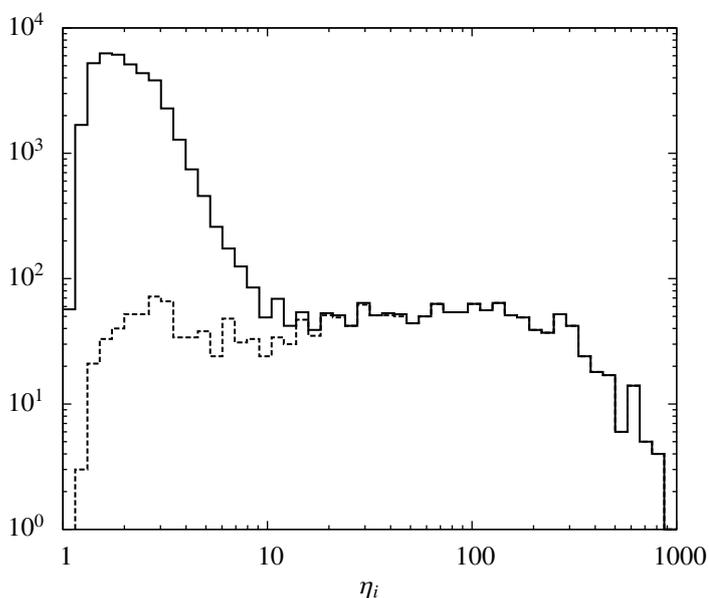}
  \caption{\label{fig:etas}Histogram of the error variance correction
    factors $\hat{\eta}_i$ for the data points of the SIP category in the simulation discussed in Sect.~\ref{sec:simulation}. The
    solid line shows the histogram for all data points in this
    category, while the dashed line shows the histogram only for the
    data points for which the noise variance was indeed increased in
    the simulation. Both axes are plotted logarithmically.}
\end{figure}

\begin{figure}
  \input{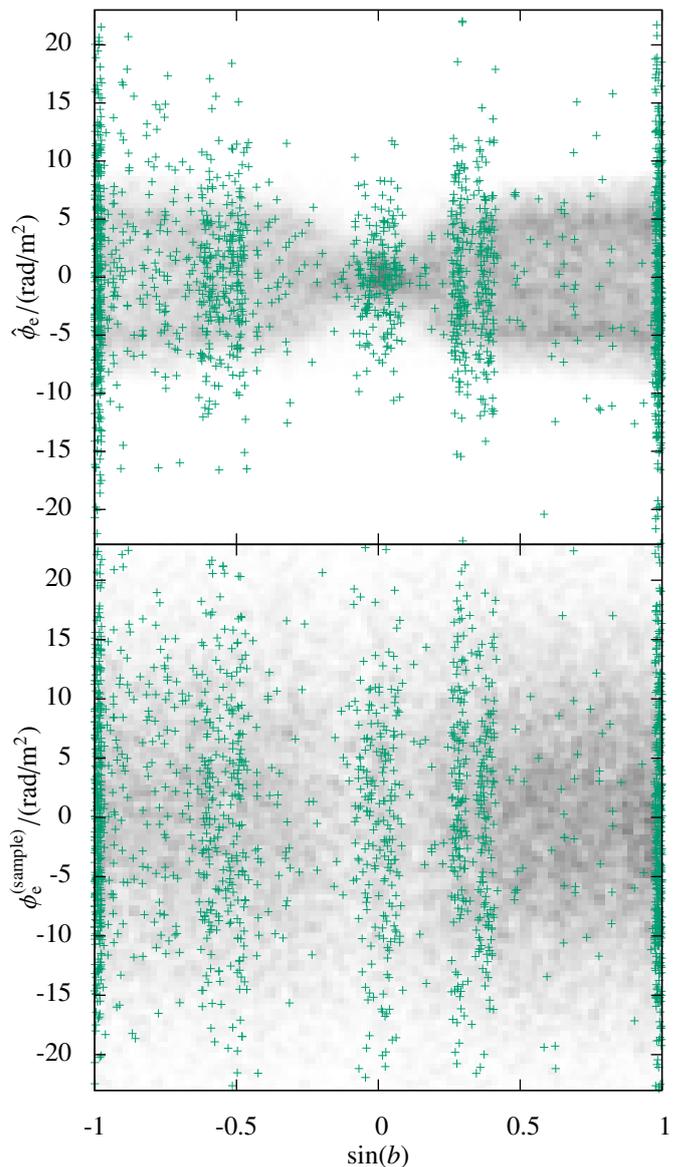}
  \caption{\label{fig:extragal}Extragalactic contribution to each source's observed Faraday depth versus Galactic latitude for the simulated scenario discussed in Sect.~\ref{sec:simulation}. The upper panel shows the approximate posterior mean estimate as calculated by our algorithm, the lower panel shows a random sample drawn from the posterior PDF. The density of data points of the SIP category, i.e., with noise variance correction factors, is plotted in grayscale, data points of the VIP category, i.e., without noise variance correction factors, are plotted as green $+$-signs.}
\end{figure}

\subsubsection{Results}

The different iterations of the Galactic latitude profile are shown in Fig.~\ref{fig:profiles}. Figure~\ref{fig:powerspectra} shows the reconstructed angular power spectra at the end of each iteration with a given latitude profile. We stop our reconstruction after six iterations.
The data contain information only on scales that are larger than the angular separation of sources. This separation is different in different regions of the sky. Typical source separations are on the order of \SI{1}{\degree}, however, the data are still dominated by noise and extragalactic contributions on this scale. This can explain the mismatch between the true and reconstructed power spectra even on scales that are a factor of a few larger than the typical source separation.

The reconstructed map of the dimensionless isotropic Gaussian random
field is shown in the left column of Fig.~\ref{fig:maps}, along with a map quantifying
its uncertainty per pixel, given by $\mathrm{diag}{(D)}$ (see Eq.~\eqref{eq:defD}). In this
simulated scenario, where we know what the map is that we are trying
to reconstruct, we can quantify this uncertainty information by
checking the number of pixels for which the true value lies within the interval
given by the reconstruction plus or minus the uncertainty, i.e., $m
\pm \sqrt{\mathrm{diag}{(D)}}$. We find in our example that this is the case
for $66\%$ of the pixels, confirming that the Gaussian approximation to the posterior PDF that we are calculating is not too far from the true posterior. The right column of Fig.~\ref{fig:maps} shows the same
for the physical Faraday depth. Sixty-three percent of the pixels have an estimated Galactic Faraday depth that lies within the approximate $1\sigma$ interval around the simulated value. These maps demonstrate that, with the data that we have simulated, the reconstruction
and the true map agree on large and intermediate scales. Only the
small-scale features are missing in the reconstructed map. This effect
is of course more prominent in regions of the sky where the data
density is lower.

To study the reconstructed error variance correction factors,
$(\hat{\eta}_i)_i$, we plot a histogram of these in
Fig.~\ref{fig:etas}. Clearly, the noise variance was increased for a
significant fraction of the data points in the SIP category. The
mean value for $\hat{\eta}_i$ is $6.2$ and the median is $2.0$. Also plotted
in Fig.~\ref{fig:etas} is a histogram only for the data points for
which the error variance was indeed increased in the simulation. The
mean and median of these $\hat{\eta}_i$-factors is $83.1$ and $32.0$,
respectively. Taking the mean and median for all data points in the
SIP category for which the error variance was \emph{not} increased
in the simulation, on the other hand, yields $2.2$ and $2.0$, respectively. So while the algorithm tends to increase all error bars slightly, there is clearly a trend  for the error bars of the right data points to be increased much more severely.

The factor $\eta_\mathrm{e}$ that corrects the assumed extragalactic variance is reconstructed in this example to be $2.6$, corresponding to a standard deviation of
\begin{equation}
	\hat{\sigma}_\mathrm{e} = \sqrt{\hat{\eta}_\mathrm{e} \, \sigma_\mathrm{e}^{2}} \approx \SI{10.6}{\radian/m^2},
\end{equation}
which is close to the value of $\sigma_\mathrm{e}^{\mathrm{(true)}} = \SI{10}{\radian/m^2}$ that we put into the simulation. Therefore we can conclude that with our algorithm we are able to reconstruct the variance of the extragalactic contribution with high precision. In principle, we could quantify the uncertainty of this estimate by taking the second derivative of the PDF given by Eq.~\eqref{eq:Poflogetae} at its maximum. However, we expect that the error in this estimate will in reality be dominated by the limitations of our assumptions, i.e., the fact that the extragalactic Faraday rotation is not exactly an isotropic, uncorrelated, Gaussian random field, and the ambiguity of the categorization of the data points, and not so much by the statistical information content of the data. We therefore believe that the quantification of the statistical uncertainty would be potentially misleading and therefore is not worth the computational effort. Even in this simulated scenario, the difference between the reconstructed \SI{10.6}{\radian/m^2} and the assumed \SI{10}{\radian/m^2} is probably mostly due to the approximations made in the derivation of our filter formulas.

Finally, in Fig.~\ref{fig:extragal}, we plot the resulting
extragalactic contributions for each data point versus Galactic
latitude. Some artifacts of the filtering procedure
are obviously present in the resulting estimate. We discuss these in the following subsection.

\subsubsection{Features in the extragalactic estimates}
\label{sec:features}

In the top panel of Fig.~\ref{fig:extragal}, we plot the estimates for the data points of the two different categories in different colors. The plot shows that the data points in the SIP category end up with estimates that have a clear latitude dependence and a rather sharp cut-off around $\left|\hat{\phi}_\mathrm{e}\right| \approx \SI{8}{\radian/m^2}$. The estimates for the data points of the VIP category also show a dependence on Galactic latitude, however, their spread is generally larger and their distribution does not exhibit a sharp cut-off.

It is important to note that the quantities plotted in the top half of Fig.~\ref{fig:extragal} are \emph{not} the extragalactic contributions, but only \emph{estimates} of these. As we pointed out earlier, given the degeneracy of the problem and the uncertainties involved, a definitive determination of the extragalactic contributions is not possible. Any separation of the observed Faraday rotation measurements into Galactic, extragalactic, and noise contributions is a trade-off between the three and therefore each of the three estimates affects the other two. Any result will have to be probabilistic in nature and based on assumptions about the properties of these contributions, no matter how sophisticated the analysis method or how good the data set. The analysis presented here does not make any prediction regarding the shape of the resulting sample distributions. Therefore, even if the reasons for the change in shape with Galactic latitude are not immediately apparent, there is also no reason to expect that the shape of the distribution of the \emph{estimate} $\hat{\phi}_\mathrm{e}$ should not change with Galactic latitude.

As we show in Appendix~\ref{app:derivation}, the posterior for the extragalactic contributions can be approximated as
\begin{equation}
	\label{eq:posteriorforextragal}
    \mathcal{P}(\phi_\mathrm{e}|d,G,E,N) = \mathcal{G}{\left(\phi_\mathrm{e} - \hat{\phi}_\mathrm{e},D_\mathrm{e} = \left(E^{-1} + \left(G + N\right)^{-1}\right)^{-1} \right)},
\end{equation}
with a Wiener filter estimate as mean and a covariance $D_\mathrm{e}$ describing the variance that is missing from the estimate itself \citep[see e.g.,][]{ensslin-2009}. Near the Galactic plane, the variance of the Galactic Faraday depth is greatly enhanced. Therefore, the entries of $G$ corresponding to lines of sight at low absolute latitudes are comparatively large and these cause $\hat{\phi}_\mathrm{e}$ plotted in the top panel of Fig.~\ref{fig:extragal} to be smaller close to the Galactic plane. However, the covariance matrix $D_\mathrm{e}$ describing the posterior Gaussian for the extragalactic contributions consequently also encodes a higher uncertainty of the estimate near the Galactic plane. So while the estimated extragalactic contributions tend to be smaller in modulus near the Galactic plane, the uncertainty of the estimate is higher.

The qualitative differences between the estimates $\hat{\phi}_\mathrm{e}$ for the VIP and SIP data categories are not surprising either, since these estimates are calculated in different ways. In essence, the estimation of the extragalactic contributions is easier when the Galactic and noise contributions are more tightly constrained. Therefore, the algorithm will find seemingly large extragalactic contributions more trustworthy for data points for which this is the case. This can likely explain the tendency for larger estimates for the data points of the VIP category, for which the noise variance cannot be increased. The sharp cut-off for the estimates for the data points of the SIP category can be interpreted as a threshold beyond which our assumptions make it more believable that the noise variance should be increased than that the extragalactic contribution is larger.

So while the absolute values of the estimates tend to be smaller for data points near the Galactic plane and for data points in the SIP category, the corresponding uncertainty is also higher. One way of looking at the estimate together with its uncertainty is to draw random realizations from the posterior PDF. Each realization represents one possible configuration that is not ruled out by the data. By drawing the realizations from the posterior, one more often draws configurations that are well supported by the data than the ones that are marginally possible. While it is true that the most probable configuration is given by the posterior maximum, $\hat{\phi}_\mathrm{e}$, this does not need to be a typical configuration in any sense. For example, while the configuration plotted in the top panel of Fig.~\ref{fig:extragal} with its small absolute values near the Galactic plane is more probable than any other specific realization, there are many more realizations with typically higher absolute values near the plane that are also not ruled out by the posterior, Eq.~\eqref{eq:posteriorforextragal}. Due to the higher posterior uncertainty near the Galactic plane, different realizations drawn from the posterior distribution will vary wildly for data points at low absolute Galactic latitude.

One such posterior sample is plotted in the lower panel of Fig.~\ref{fig:extragal}. Obviously, the artifacts visible in the estimate of the extragalactic contributions are mostly compensated by the uncertainties of these estimates. As is true for any scientific analysis, our state of knowledge about the extragalactic contributions to Faraday rotation after analyzing the data set cannot be described completely by an estimate, but only by a probability distribution. The random sample drawn from this distribution shows that this distribution does not exhibit any crude artifacts of the analysis, only the attempted summary in a single estimate does.

\section{Application to real data}
\label{sec:realworld}

\subsection{Description of the data}
\label{sec:datadescription}

In the following, we make use of the data catalogs assembled by \citet{oppermann-2012} and described in their Table~1. We add the new catalog of \citet{mao-2012}, which has the same observing specifications as the catalog of \citet{vaneck-2011}, detailed in the table, except for the number of sources and their locations. Altogether, this data set consists of $41\,632$ observationally estimated Faraday depths for extragalactic sources. The extragalactic nature of the sources is not entirely guaranteed. While it is possible that a few of the data points correspond to pulsars in the Milky Way, we note that the overwhelming majority of the sources has to be extragalactic. Pulsars, for which not the complete line of sight to the outer edge of the Milky Way is probed by the observations, provide one more reason to attempt a reconstruction that is robust against an incomplete description of the observational uncertainties.

The data set is rather inhomogeneous both spatially, with a relatively sparse source population in the southern equatorial hemisphere, and with respect to the observational parameters, ranging from linear fits to polarization angle measurements in two adjacent frequency bands to RM synthesis studies over wide bands in $\lambda^2$-space.

In the following, we multiply the published error bars of \citet{taylor-2009} by a factor $1.22$, according to Sect.~4.2.1 of \citet{stil-2011}.

\subsection{Possible splits}
\label{sec:possiblesplits}

The algorithm presented in the previous section hinges on the assumption that we can split the data set into a subset for which the likelihood is fully described by the published Gaussian error bars (VIP data) and another subset for which this is not necessarily the case (SIP data). How to judge whether a data point should be in the VIP category or the SIP category is not clear. Aspects that are to be considered in this decision are that a continuous frequency coverage eliminates the risk of a polarization angle rotation by a multiple of $\pi$ between bands and that a wider coverage in $\lambda^2$-space leads to a higher resolution in Faraday depth space and therefore a lower risk of misleading results occurring from several emission components within the same beam, as described by \citet{farnsworth-2011} and \citet{kumazaki-2014}. This demands a large fractional band-width. Furthermore, the estimation of the variance of the extragalactic contribution relies mostly on the data points that we assign to the VIP category. As we pointed out earlier, this estimation is complicated by large contributions from the Milky Way and large noise contributions. It is therefore desirable to have at least some data points of the VIP category away from the Galactic plane. Finally, it is good to split the data in a conservative way. Assigning to the SIP category a data point for which the likelihood is well described by the given error bar will not bias the result, only increase the posterior uncertainty. On the other hand, assigning a data point for which the likelihood is insufficiently known to the VIP category will in most cases lead to an overestimated variance of the extragalactic contribution and thus influence all other results of the reconstruction.

\begin{figure*}
	\includegraphics[width=0.5\textwidth]{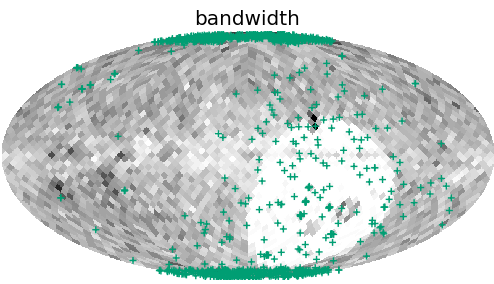}
	\includegraphics[width=0.5\textwidth]{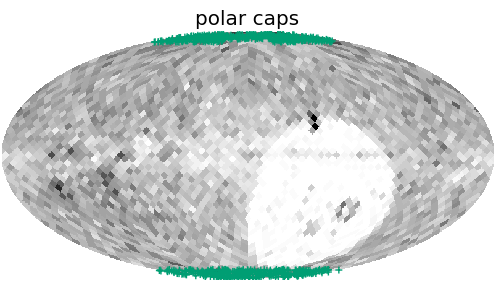}
	\includegraphics[width=0.5\textwidth]{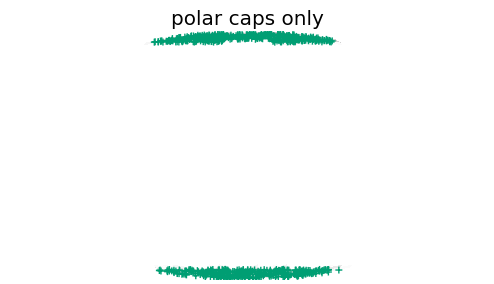}
	\includegraphics[width=0.5\textwidth]{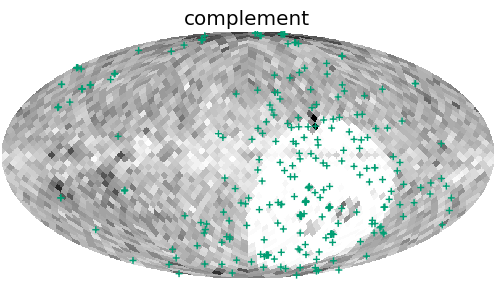}
	\includegraphics[width=0.5\textwidth]{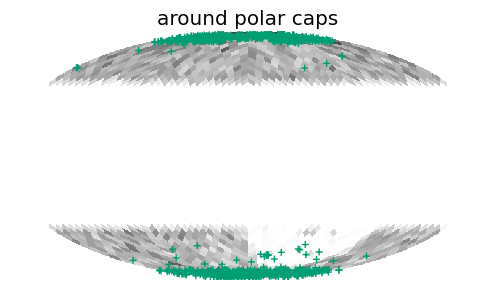}
	\includegraphics[width=0.5\textwidth]{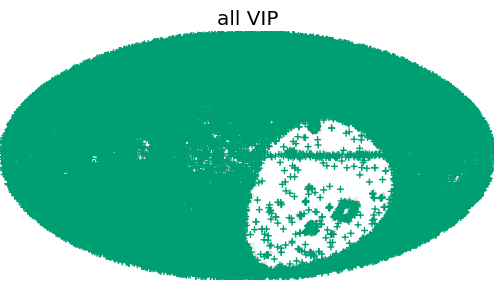}
	\caption{\label{fig:splits_dataplot}Locations of the data points in the sky for the different data splits. The grayscale shows the density of all data points considered, while VIP data points are marked by the green $+$-signs. The labels refer to the first six data splits described in Sect.~\ref{sec:possiblesplits} and in Table~\ref{tab:sigma_e}.}
\end{figure*}

Instead of arguing for a single definitive split, we will explore a set of different possibilities. This will enable us to make statements about the reliability of the results. The following ways of splitting the data set will be used:
\begin{enumerate}
	\item Five catalogs with a wide frequency coverage at particularly low frequencies are regarded as data of the VIP category. These are the catalogs referred to as O'Sullivan (O'Sullivan, private communication, 2011), Heald \citep{heald-2009}, Schnitzeler (Schnitzeler, private communication, 2011), as well as Mao SouthCap and Mao NorthCap \citep{mao-2010} in Table~1 of \citet{oppermann-2012}. All other data are considered part of the SIP category. We will refer to this split as the `bandwidth' split.
	\item Only the two catalogs consisting entirely of data points near the Galactic poles, i.e., Mao SouthCap and Mao NorthCap, are considered as the VIP category. These combine the demand for large coverage in $\lambda^2$-space with a low foreground region in the sky. All other data are considered SIP category data. We will refer to this split as the `polar caps' split.
	\item Only the Mao NorthCap and Mao SouthCap data are used. These are considered data of the VIP category. All other data are completely ignored, i.e., there is no SIP category of data. This means that the reconstruction is completely insensitive to anything that happens at low Galactic latitudes and therefore our assumption of approximate isotropy for the Galactic foreground can be expected to be rather accurate in this case. We will refer to this split as `polar caps only'.
	\item The data of the O'Sullivan, Heald, and Schnitzeler catalogs are regarded as being of the VIP category, all other data are in the SIP category. This is done to see the effect that having VIP data in regions with a large foreground may have on the result. We will refer to this split as the `complement' split, as it regards as VIP data the points that are regarded as VIP under the `bandwidth' condition but not under the `polar caps' condition.
	\item Only data points with Galactic latitudes that satisfy $\left|b\right|>\SI{45}{\degree}$ are considered at all. Of these, the data that stem from any of the O'Sullivan, Heald, Mao NorthCap, Mao SouthCap, or Schnitzeler catalogs and additionally satisfy $\left|b\right|>\SI{55}{\degree}$ are considered as comprising the VIP category. The last condition is introduced to avoid any potential boundary effects on the reconstruction of the extragalactic variance. Otherwise, this is essentially an extension of the `polar caps only' ansatz, which adds a few data points of the VIP category and a significant number of data points in the SIP category. We will refer to this split as the `around polar caps' split.
	\item All data points are considered part of the VIP category. The SIP category is empty, i.e., the observational uncertainty is regarded as precisely reliable for each and every data point. We regard this split as a cross-check to see whether the algorithm behaves in the expected way if we contaminate the VIP data category with data points for which the uncertainty information is incomplete. We will refer to this assumption as `all VIP'.
	\item Only the data from the Mao NorthCap catalog are considered part of the VIP category. All other data points are considered SIP category data. We will refer to this split as `north polar'.
	\item Only the data from the Mao SouthCap catalog are considered part of the VIP category. All other data points are considered SIP category data. We will use this split and the one before as consistency checks for the results of the `polar caps' split. This split will be referred to as `south polar'.
	\item 10\,000 randomly chosen data points are assigned to the VIP data category. The rest of the data (i.e., 31632 measurements) are used as SIP category data. We will refer to this as the `random' split.
\end{enumerate}
Table \ref{tab:sigma_e} gives an overview of the data splits we consider and Fig.~\ref{fig:splits_dataplot} shows the locations of the VIP and SIP data points in the first six splits. The majority of the data points (37\,543 points) stems from the NVSS RM catalog \citep{taylor-2009}. These are either regarded as part of the SIP category or removed completely in our splits, except for the `all VIP' and `random' cases.

\subsection{Results and discussion}

\renewcommand{\arraystretch}{1.1}
\begin{table*}
	\caption{\label{tab:sigma_e}Overview of the different data splits considered for the analysis of the observational data. The first column gives the name by which the split is referred to, the second one describes the criterion by which data are assigned to the VIP category, the third column lists the data catalogs whose data points are assigned to the VIP category, the fourth column gives the criterion for a data point to be part of the SIP category. The resulting estimate for the dispersion of the extragalactic contributions is given in the second to last column, calculated from the correction factor given in the third to last column, and the last column lists the figures that show results obtained under the data split in question.}
	\centering
	\begin{tabular}{l m{3cm} m{3.5cm} m{3cm} r @{.}l r @{.}l m{1.2cm}}
		\hline\hline
		data split & condition for VIP data & VIP catalogs & condition for SIP data & \multicolumn{2}{c}{$\hat{\eta}_\mathrm{e}$} &\multicolumn{2}{c}{$\hat{\sigma}_\mathrm{e}/(\SI{}{\radian/m^2})$} & Figs.\\
		\hline
		`bandwidth' & large bandwidth & O'Sullivan\tablefootmark{a}, Heald\tablefootmark{b}, Schnitzeler\tablefootmark{c}, Mao\ SouthCap\tablefootmark{d}, Mao~NorthCap\tablefootmark{d} & remaining data & 1 & 57 & 8 & 3 & \ref{fig:splits_dataplot} - \ref{fig:eta_i_comparison}\\
		`polar caps' & large bandwidth \&~low foreground & Mao~SouthCap\tablefootmark{d}, Mao~NorthCap\tablefootmark{d} & remaining data & 1 & 09 & 6 & 9 & \ref{fig:splits_dataplot} - \ref{fig:maps-polarcaps}, \ref{fig:posteriorstddev} - \ref{fig:posteriorcorrelations}\\
		`polar caps only' & large bandwidth \&~low foreground & Mao~SouthCap\tablefootmark{d}, Mao~NorthCap\tablefootmark{d} & - & 1 & 12 & 7 & 0 & \ref{fig:splits_dataplot} - \ref{fig:m_e_comparison}\\
		`complement' & large bandwidth \&~not low foreground & O'Sullivan\tablefootmark{a}, Heald\tablefootmark{b}, Schnitzeler\tablefootmark{c} & remaining data & 7 & 86 & 18 & 5 & \ref{fig:splits_dataplot} - \ref{fig:eta_i_comparison}\\
		`around polar caps' & large bandwidth \&~$|b|>\SI{55}{\degree}$ & Mao~SouthCap\tablefootmark{d}, Mao~NorthCap\tablefootmark{d}, part of O'Sullivan\tablefootmark{a}, Heald\tablefootmark{b}, Schnitzeler\tablefootmark{c} & $|b|>\SI{45}{\degree}$ & 1 & 31 & 7 & 5 & \ref{fig:splits_dataplot} - \ref{fig:eta_i_comparison}\\
		`all VIP' & all data & all & - & 25 & 39 & 33 & 3 & \ref{fig:splits_dataplot} - \ref{fig:m_e_comparison}\\
		`random' & 10\,000 random data points & from all catalogs & remaining data & 18 & 45 & 28 & 4 & - \\
		`north polar' & large bandwidth \&~near Galactic north pole & Mao~NorthCap\tablefootmark{d} & remaining data & 1 & 01 & 6 & 6 & - \\
		`south polar' & large bandwidth \&~near Galactic south pole & Mao~SouthCap\tablefootmark{d} & remaining data & 1 & 19 & 7 & 2 & - \\
		\hline
	\end{tabular}
	\tablefoot{
		\tablefoottext{a}{O'Sullivan, private communication (2011)} \tablefoottext{b}{\citet{heald-2009}} \tablefoottext{c}{Schnitzeler, private communication (2011)} \tablefoottext{d}{\citet{mao-2010}}
		}
\end{table*}

\begin{figure*}
	\includegraphics[width=1.0\textwidth]{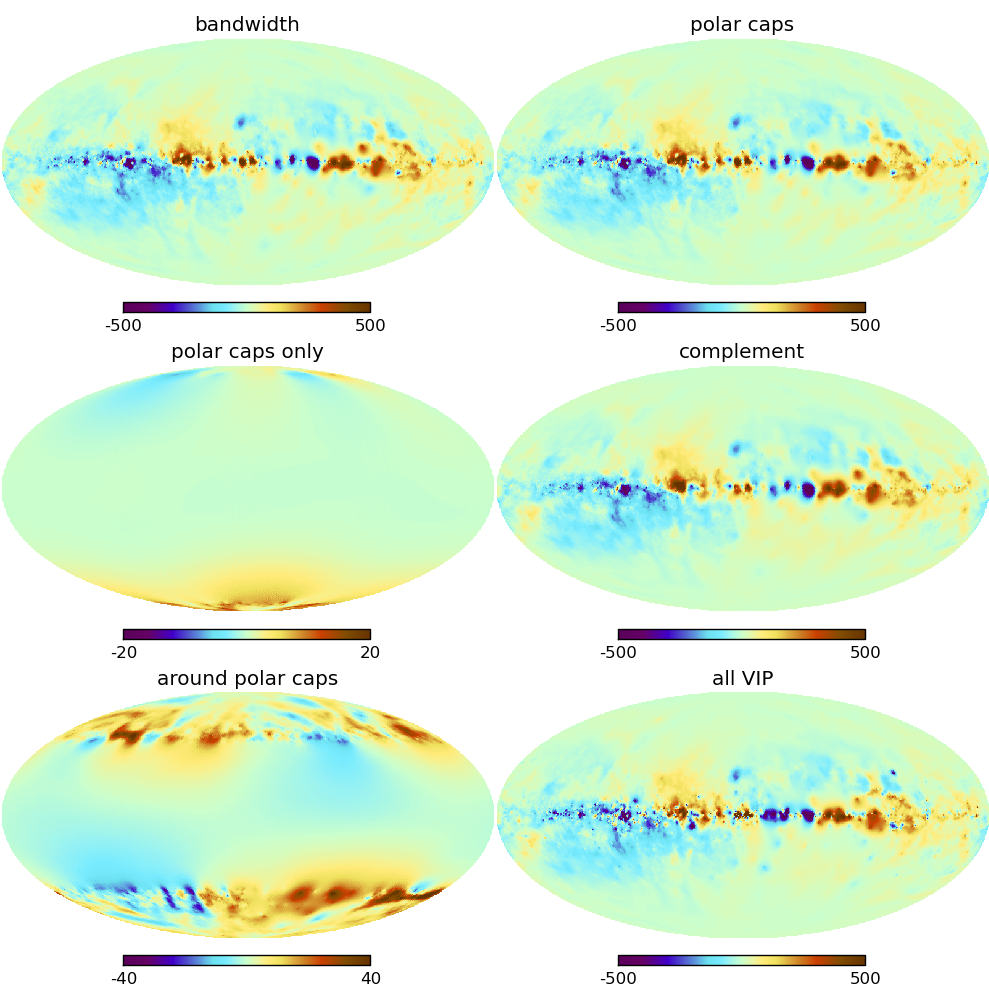}
	\caption{\label{fig:phi_maps}Reconstructions of the Galactic foreground in the first six scenarios. The labels refer to the first six data splits described in Sect.~\ref{sec:possiblesplits} and Table~\ref{tab:sigma_e}. The units are \SI{}{\radian/m^2}. The different color scales for the `polar caps only' and `around polar caps' splits differ from the ones used in the other panels.}
\end{figure*}

\begin{figure}
	\input{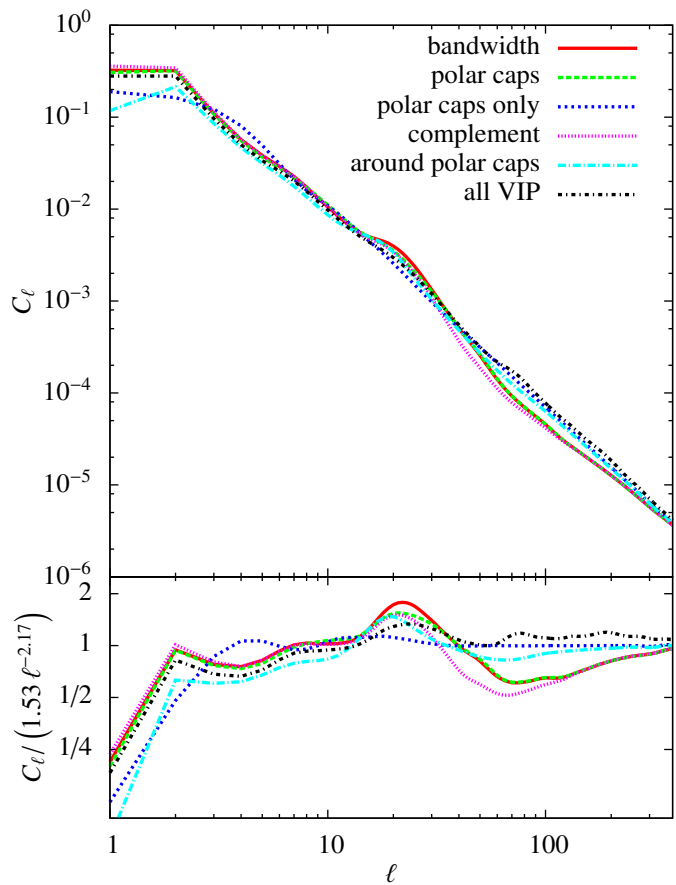}
	\caption{\label{fig:angularspectra}Reconstructed angular power spectra of the dimensionless Galactic signal field. The lower panel shows the ratio of the reconstructed spectra and a power law fit, $C_\ell = 1.53\, \ell^{-2.17}$.}
\end{figure}

\begin{figure}
	\input{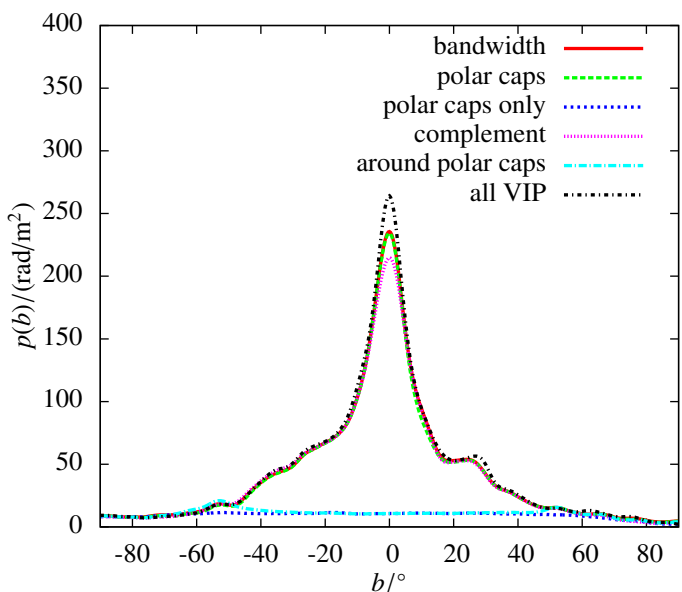}
	\caption{\label{fig:profilesreal}Reconstructed Galactic latitude profiles, $\hat{p}{(b)}$, describing the relationship between the dimensionless Galactic signal field $s$ and the physical Galactic Faraday depth $\phi_\mathrm{g}$, for the six different data splits.}
\end{figure}

\begin{figure*}
	\input{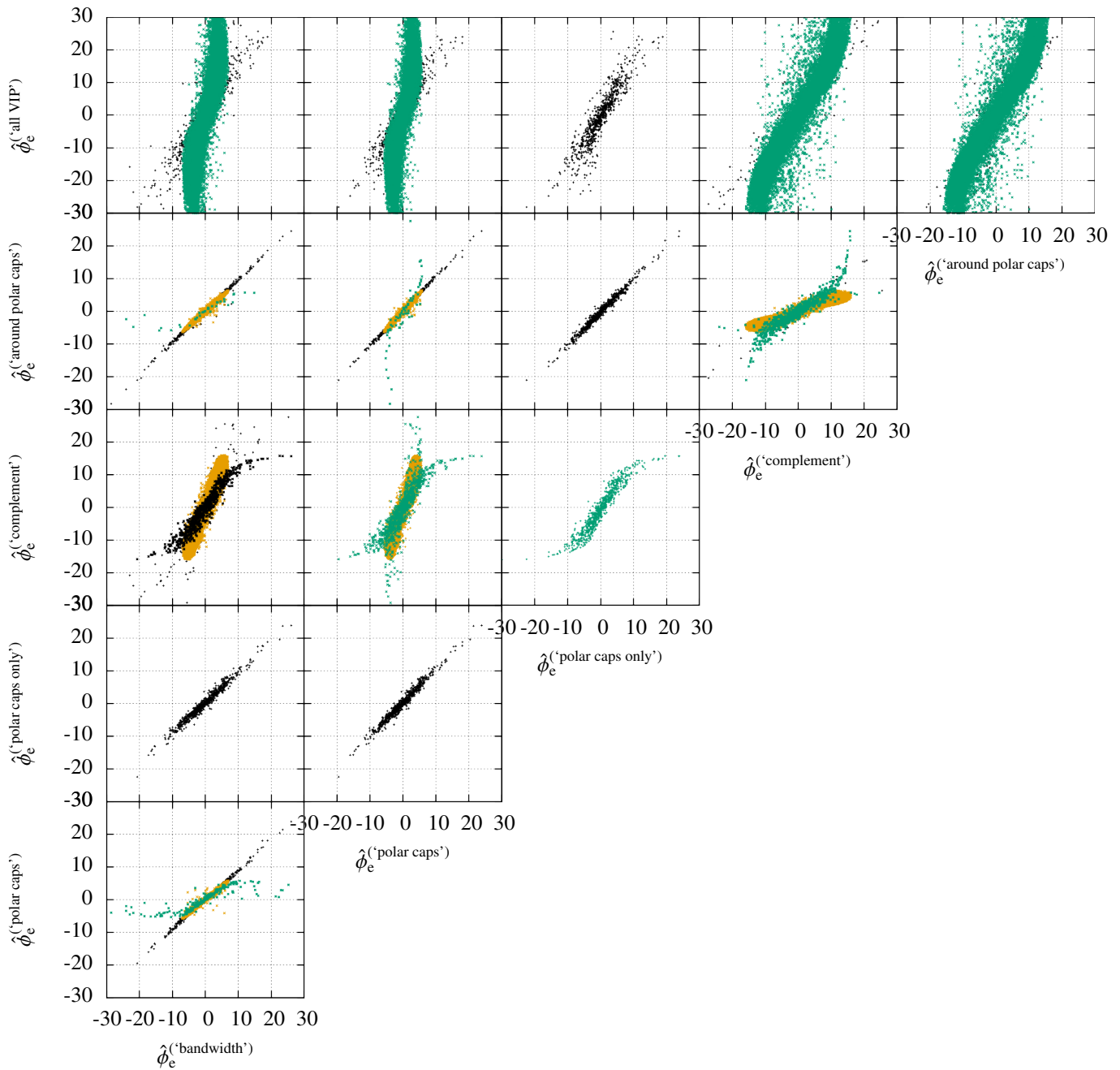}
	\caption{\label{fig:m_e_comparison}Comparison of the estimates for the extragalactic contributions, $\hat{\phi}_\mathrm{e}$, for the six different sets of assumptions described in the text. Black points are for data points that are in the VIP category under both assumptions that are being compared, green data points are in the SIP category under one of the assumptions and in the VIP category under the other, and orange points are in the SIP category under both assumptions. The units of all axes are \SI{}{\radian/m^2}.}
\end{figure*}

\begin{figure*}
	\input{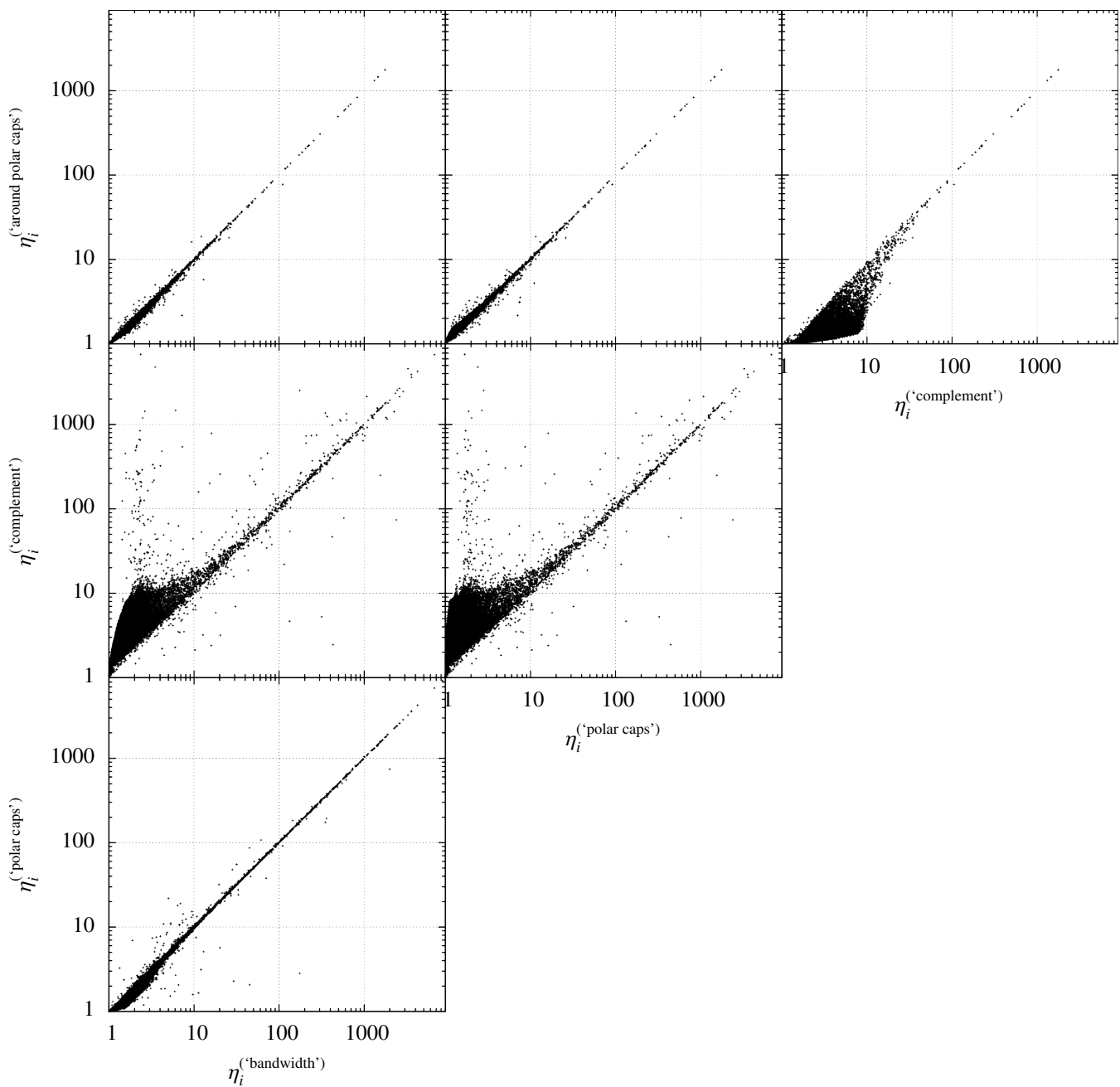}
	\caption{\label{fig:eta_i_comparison}Comparison of the error variance correction factors for the data of the SIP category under the four splits of the six under consideration that have data points of the SIP category.}
\end{figure*}

In this section, we will first discuss the results of the first six splits in detail in Sect.~\ref{sec:firstsixsplits} and then introduce the most important aspects of the remaining splits. We will argue for adopting the `polar caps' split as a reasonable fiducial model and use the `north polar' and `south polar' splits as cross-checks for the reliability of the results derived under this split in Sect.~\ref{sec:crosschecks}. Finally, we will present detailed results for the `polar caps' split in Sect.~\ref{sec:polarcapsresults}.

\subsubsection{The first six data splits}
\label{sec:firstsixsplits}

Figure~\ref{fig:phi_maps} shows the reconstructions of the Galactic contribution in the first six cases. Naturally, the reconstruction in the `polar caps only' and `around polar caps' cases suffers from the scarcity of data. The `all VIP' reconstruction shows small-scale structure, especially in the plane (for example around the Galactic center), that is washed out in the other reconstructions, again as expected. The three other reconstructions that make use of the entirety of the data are rather similar. Some details, however, do differ. We note for example the blob of positive Galactic Faraday depth at $l\approx\SI{275}{\degree}$, $b\approx\SI{10}{\degree}$ that is present in the `bandwidth' and `complement' reconstructions but not in the `polar caps' reconstruction. The reasons for these differences are not immediately apparent, and are most likely due to the interplay of all involved quantities and possibly an instability with respect to numerical inaccuracies.

The reconstructed angular power spectra of the dimensionless Galactic signal fields in the first six cases are shown in Fig.~\ref{fig:angularspectra}. Evidently, the resulting spectra are all very similar. In the `polar caps only' and `around polar caps' cases, in which a large fraction of the data were ignored, the result is closer to a pure power law, since the spectral smoothness prior becomes more important in these cases. A power law with a spectral index of $-2.17$ is a good fit to these spectra, as was already found by \citet{oppermann-2012}.

Figure~\ref{fig:profilesreal} shows the variance profiles for the Galactic contribution introduced in Eq.~\eqref{eq:intro_p} that result from the six different data splits. Obviously, in the `polar caps only' and `around polar caps' cases, the profile function is only reconstructed well near the poles, as all the low latitude data are ignored. Among the other reconstructions, the profile functions do not differ greatly, with the main differences appearing near the Galactic plane. The `all VIP' ansatz leads to a higher variance near the Galactic plane, whereas a smaller fraction of VIP data leads to a more heavily smoothed Galactic reconstruction and therefore less variance and a slightly lower profile function, as exemplified by the `polar caps' and `complement' splits.

The reconstructed values, $\hat{\sigma}_\mathrm{e}$, of the extragalactic dispersion are presented in Table~\ref{tab:sigma_e}. We note that the `polar caps', `polar caps only', and `around polar caps' numbers are rather similar. For all of these reconstructions, the VIP category of data is dominated by the Mao NorthCap and Mao SouthCap catalogs. The number provided by the `bandwidth' reconstruction is not very different either. In this case the Mao SouthCap and Mao NorthCap catalogs still comprise more than $70\%$ of the VIP data category. The `complement' number differs significantly, indicating that the assumptions made for the VIP category of data points are probably not met by all of the data points in the O'Sullivan, Heald, and Schnitzeler catalogs. Another factor here may be the fact that in the `complement' split, the data of the VIP category are few and far between (a total of 281 data points, mostly in the southern equatorial hemisphere, some in the northern hemisphere). Obviously, the `all VIP' scenario yields a number that is even more in disagreement.

Figure~\ref{fig:m_e_comparison} shows a comparison of the estimates for the extragalactic contributions under the six different assumptions. Black points are for data points that are in the VIP category under both assumptions that are being compared in each individual panel. As can be seen in the figure, the estimates for the extragalactic contributions are basically the same for data points that are in the VIP category for both compared scenarios if the estimate $\hat{\sigma}_\mathrm{e}$ is similar in the two scenarios. A significantly larger estimate $\hat{\sigma}_\mathrm{e}$, however, leads to increased estimates $\hat{\phi}_\mathrm{e}$ as well, as can be seen most clearly in the top row of the figure. The black dots in these panels still follow a linear relationship, but the slope deviates from one. These trends are a direct consequence of Eq.~\eqref{eq:exest}, which means that, after the correction factors $\hat{\eta}_\mathrm{e}$ and $\hat{\eta}_i$ are fixed, the difference between the observed value and the estimated Galactic contribution is split between the estimates of the extragalactic and noise contributions according to their expected variances, i.e.,
\begin{equation}
	\label{eq:split1}
	\hat{\phi}_{\mathrm{e},i} = \frac{\hat{\sigma}_\mathrm{e}^2}{\hat{\eta}_i\left(\sigma_\mathrm{e}^2 + \sigma_i^2\right)} \left( d_i - \hat{\phi}_{\mathrm{g},i} \right)
\end{equation}
for SIP data and
\begin{equation}
	\label{eq:split2}
	\hat{\phi}_{\mathrm{e},i} = \frac{\hat{\sigma}_\mathrm{e}^2}{\hat{\sigma}_\mathrm{e}^2 + \sigma_i^2} \left( d_i - \hat{\phi}_{\mathrm{g},i} \right)
\end{equation}
for VIP data. Some consequences of these equations are discussed in Appendix~\ref{app:bryan}.

Orange points in Fig.~\ref{fig:m_e_comparison} are data points that are in the SIP category under both of the two assumptions that are being compared. These orange dots show essentially the same effect as the black dots, namely that the estimates are the same if the estimate, $\hat{\sigma}_\mathrm{e}$, of the extragalactic dispersion is the same and a higher estimate $\hat{\sigma}_\mathrm{e}$ results in a higher estimate $\hat{\phi}_\mathrm{e}$. The latter effect is visible whenever the `complement' estimate is part of the comparison. Overall, the orange dots have a smaller dispersion than the black ones. This is expected, as part of the discrepancy between data and Galactic reconstruction can be explained by increased error bars in case of data in the SIP category.

Finally, the green points in Fig.~\ref{fig:m_e_comparison} show data points that are in the SIP category under one of the assumptions that are being compared and in the VIP category under the other. These points in the figure show that allowing the noise variance to be corrected upward for a data point will keep the estimate of its extragalactic contribution small. We note the bifurcation in the green points in the comparisons between the `polar caps' and `complement' estimates and between the `around polar caps' and `complement' estimates. This is due to data points that are in the SIP category under the `complement' split on the one hand and data points being in the SIP category under the `polar caps' and `around polar caps' splits on the other hand. Overall, all the trends exhibited by the estimates shown in Fig.~\ref{fig:m_e_comparison} follow the expectation.

The effect of interpreting the error bars as a complete description of the likelihood functions can be seen clearly in the first row of panels in Fig.~\ref{fig:m_e_comparison}. For many data points the observed Faraday depth cannot be explained by the Galactic foreground reconstruction and the published noise variance alone. The algorithm will therefore increase the dispersion $\hat{\sigma}_\mathrm{e}$ of the extragalactic contributions until it agrees with the dispersion of the remaining differences. This leads to the high reconstructed value $\hat{\sigma}_\mathrm{e}$ and the large estimates $\hat{\phi}_\mathrm{e}$ seen in Fig.~\ref{fig:m_e_comparison} in the `all VIP' case. If, on the other hand, only a subset of the error bars are assumed to accurately describe the likelihood functions, the estimate $\hat{\sigma}_\mathrm{e}$ will be dominated by this subset of the data points (i.e., the VIP data). Furthermore, if this assumption is indeed true for the chosen subset, the estimate $\hat{\sigma}_{\mathrm{e}}$ will naturally be lower. Consequently, the algorithm will explain large differences between observed Faraday depths for the SIP data and the Galactic foreground reconstruction as largely due to a likelihood function that is wider than described by the error bar, i.e., an error variance correction factor $\hat{\eta}_i$ that is significantly larger than one. This keeps the estimated extragalactic contributions relatively small, as plotted on the horizontal axes of the first row of panels in Fig.~\ref{fig:m_e_comparison}.

Figure~\ref{fig:eta_i_comparison} compares the error variance correction factors, $\hat{\eta}_i$, for data that are in the SIP category under two sets of assumptions. This comparison can only be done for the four splits of the six under consideration that do have data of the SIP category, namely the `bandwidth', `polar caps', `complement', and `around polar caps' splits. This shows that the $\hat{\eta}_i$ factors are almost unaffected by the differences between the `bandwidth', `polar caps', and `around polar caps' assumptions. The `complement' split, however, leads to systematically larger correction factors. This is most likely due to a contamination of the VIP data category by data for which the likelihood is not well described by the observational error bars. Such a data point in the VIP category will lead to an increase of the estimated variance of the extragalactic contribution. However, the data points that were appropriately placed into the VIP category will prevent this estimate from growing enough to completely explain the faulty data point. As a result, the faulty data point will still have a strong influence on the reconstruction of the Galactic foreground in its vicinity, thus necessitating increased noise variances for other data points to be consistent with this reconstruction. To check whether this is indeed a generic effect of a contaminated VIP data category, we now study the `random' split, for which we randomly assign 10\,000 of the 41\,632 data points to the VIP category and run our reconstruction algorithm under these assumptions. We find as expected a high estimate for the extragalactic dispersion of $\hat{\sigma}_\mathrm{e} = \SI{28.4}{\radian/m^2}$. Furthermore, we find the same trend for increased error variance factors as for the `complement' split. We therefore conclude that these enlarged error variances are indeed a sign for a contamination of the VIP data category with data that should have been put into the SIP category.

\subsubsection{Cross-checks for the `polar caps' split}
\label{sec:crosschecks}

From the splits that we have studied so far, we regard as most reliable the ones that use only the Mao SouthCap and Mao NorthCap data as data of the VIP category, i.e., the `polar caps' and the `polar caps only' splits. To check whether this is indeed the case or whether the values for $\hat{\sigma}_\mathrm{e}$ calculated under the splits that are dominated by these data sets are only similar by chance, we consider two additional subsets of the data for the VIP category. We perform reconstructions using the entirety of the data with only the data from the Mao NorthCap catalog and only the data from the Mao SouthCap catalog assigned to the VIP data category, the `north polar' and `south polar' splits mentioned before. The resulting values for the estimated extragalactic dispersion are $\hat{\sigma}_\mathrm{e} = \SI{6.6}{\radian/m^2}$ and $\hat{\sigma}_\mathrm{e} = \SI{7.2}{\radian/m^2}$, respectively, close to the value found in the `polar caps' split. We therefore conclude that the value for the extragalactic dispersion in reality is likely to lie somewhere in the interval between $\SI{6.0}{\radian/m^2}$ and $\SI{7.1}{\radian/m^2}$, as spanned by the results of the different splits using the Mao NorthCap and Mao SouthCap data as the only ones for which the likelihood is assumed to be precisely described by the published Gaussian error bars. It should also be noted that the extragalactic dispersion estimated in this way will at least partially include any contribution from the Milky Way that is uncorrelated on the scales probed by current observations, i.e., typically around \SI{1}{\degree}. Therefore, the purely extragalactic dispersion may be even lower.

\subsubsection{Results for the `polar caps' split}
\label{sec:polarcapsresults}

\begin{figure}
	\input{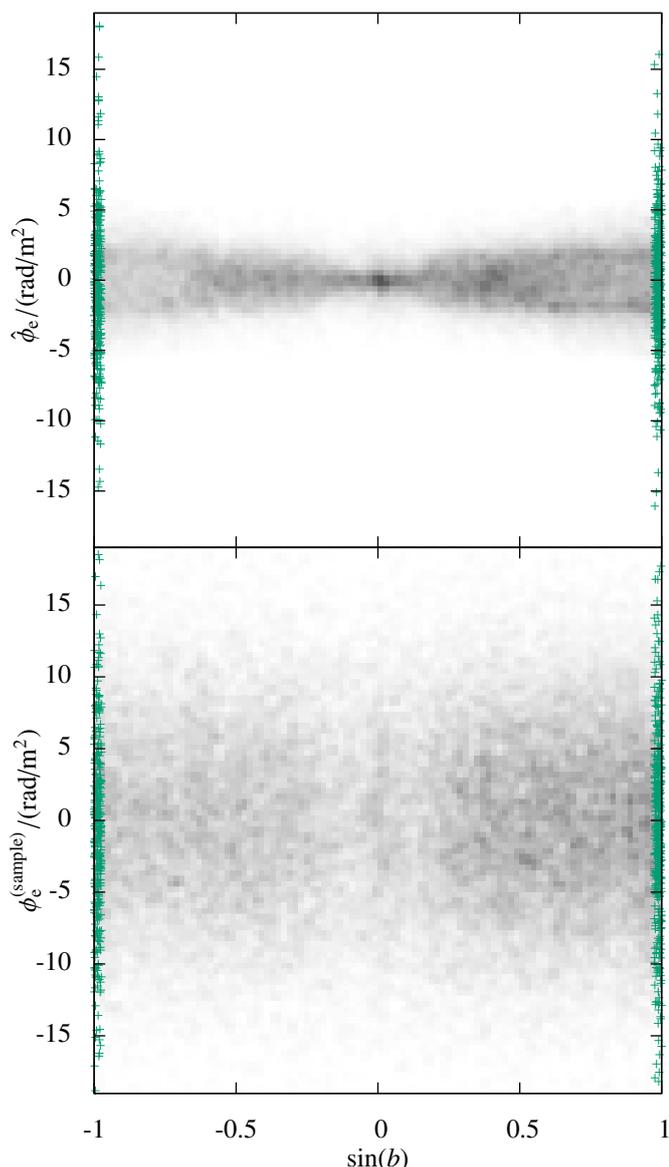}
	\caption{\label{fig:exoflat_polarcaps}\emph{Top}: Estimates for the extragalactic contribution to each source's observed Faraday depth for the `polar caps' analysis versus Galactic latitude.  \emph{Bottom}: The same for a random sample drawn from the posterior PDF for the extragalactic contributions around the mean plotted in the upper panels. In both panels, the grayscale shows the density of the data points of the SIP category, i.e., with noise variance correction factors, and data points of the VIP category, i.e., without noise variance correction factors, are plotted as green $+$-symbols.}
\end{figure}

\begin{figure*}
	\includegraphics[width=\textwidth]{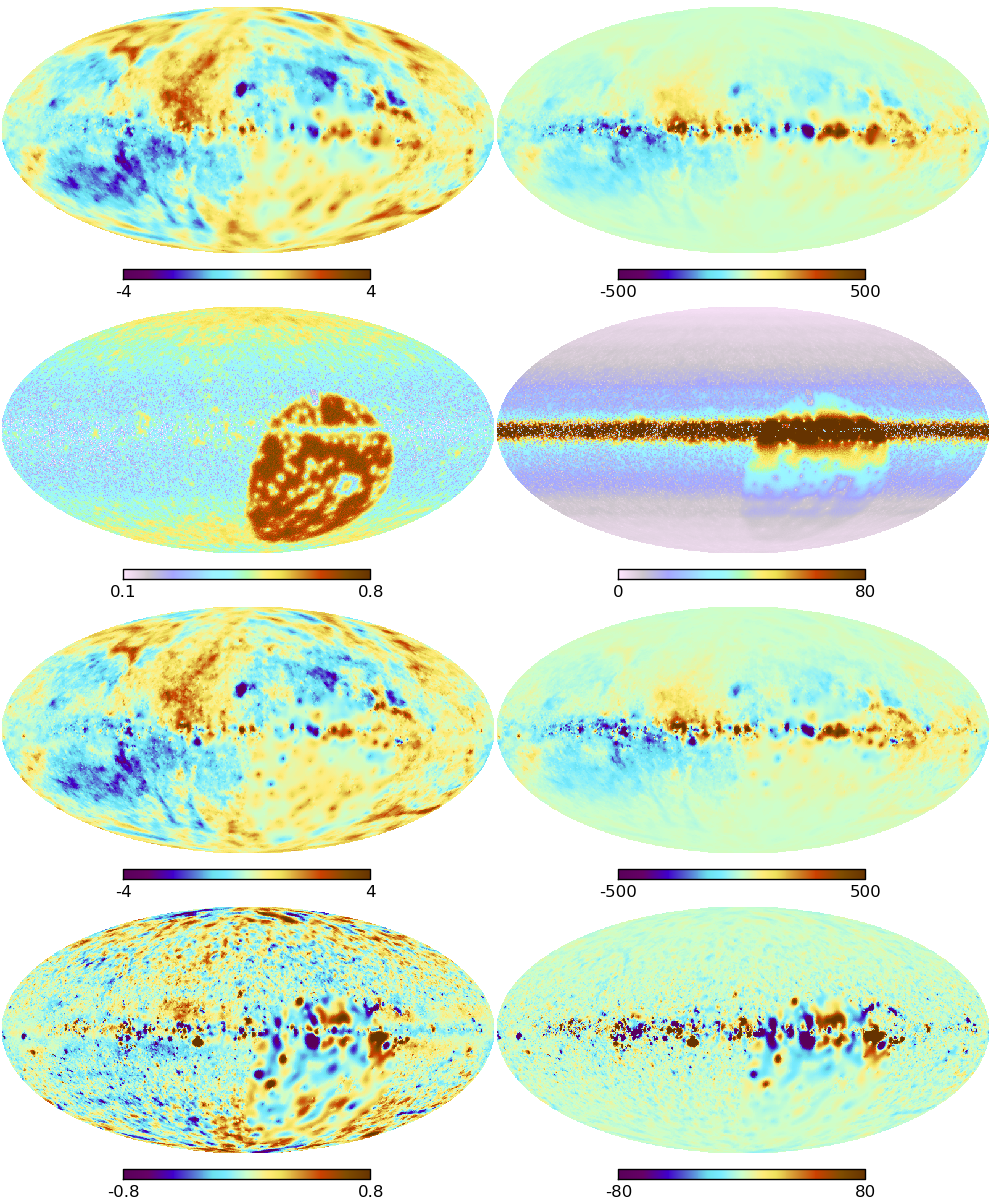}
	\caption{\label{fig:maps-polarcaps}Estimates of the Galactic contribution under the `polar caps' data split and comparison to the results of \citet{oppermann-2012}. The left column illustrates the dimensionless Galactic signal field $s$, the right column the physical Galactic Faraday depth $\phi_\mathrm{g}$ in units of $\SI{}{\radian/m^2}$. The top row shows the posterior mean estimates derived with the `polar caps' split, the second row shows the pixel-wise uncertainty of this estimate, the third row shows the result of \citet{oppermann-2012}, and the bottom row shows the result of subtracting the third row from the top row.}
\end{figure*}

We therefore adopt the `polar caps' split as our fiducial model, i.e., in the following we work with the estimate $\hat{\sigma}_\mathrm{e} = \SI{6.9}{\radian/m^2}$, which resulted from this split. Figure~\ref{fig:exoflat_polarcaps} shows the estimated extragalactic contribution for this reconstruction as a function of latitude, as well as a random sample drawn from the posterior PDF for this quantity. We note that again the distribution of the posterior mean estimates shows a latitude dependence, as was the case for the results in our simulated scenario shown in Fig.~\ref{fig:extragal}. This is due to the latitude dependence of the Galactic contribution, as discussed in Sect.~\ref{sec:features}. However, as pointed out before, this is only an \emph{estimate} for the extragalactic contributions and as such should always be regarded together with its uncertainty. To demonstrate this, we draw a random sample from the posterior PDF for the extragalactic contributions and show it in the lower panel of Fig.~\ref{fig:exoflat_polarcaps}. The sample includes the spread that is due to the remaining uncertainty of the estimate. This is larger near the Galactic plane and mostly lets the latitude dependence vanish in the bottom panel of the figure. We note, however, that there is still a slight difference between the distributions for the VIP and SIP categories. We have created a website\footnote{See \url{http://www.mpa-garching.mpg.de/ift/faraday/}} where we provide such posterior samples for use in studies of extragalactic Faraday rotation along with the other results of the `polar caps' reconstruction. The samples are again not drawn from the full posterior PDF, but from the approximate one we use in our derivations. This approximate posterior is effectively the posterior for the extragalactic contributions after fixing their variance, the error variance correction factors, the angular power spectrum of the Galactic contribution, and the Galactic latitude profile, i.e., the uncertainty due to the uncertainty of these reconstructed quantities is not represented by the samples. See Appendix~\ref{app:website} for details on the provided files and the usage of the samples, as well as a discussion of numerical artifacts present in the samples.

Finally, we show in Fig.~\ref{fig:maps-polarcaps} the results for the Galactic reconstruction using the `polar caps' data split, i.e., the reconstructed dimensionless signal field and physical Galactic Faraday depth, as well as their pixel-wise uncertainties, given by the square-root of the diagonal of the posterior covariance in position space representation. Figure~\ref{fig:maps-polarcaps} also shows a comparison to the results of \citet{oppermann-2012}. As can be seen, differences are most pronounced near the Galactic plane, in the region of scarce data in the southern equatorial hemisphere, and near the poles. These differences are due to the differences of the reconstruction algorithms. The biggest difference is of course our  special treatment of a subset of the data that will influence the Galactic reconstruction as well. Additionally, the spectral smoothness prior that we used here prevents the angular power spectrum from dropping off steeply at the smallest scales, as was found by \citet{oppermann-2012}, and thus will lead to more small-scale fluctuations in the Galactic map. Finally, we no longer allow the error variance correction factors to become smaller than one. This will in general have a suppressing effect on the small-scale structure and seems to dominate over the opposing effect, as we see more small-scale structure in the third row of the figure than in the first. While the bottom panels in Fig.~\ref{fig:maps-polarcaps} highlight the differences with respect to the old reconstruction, it should be noted that overall the differences are small. The estimate of the Galactic Faraday depth in the `polar caps' scenario lies within the uncertainty range published by \citet{oppermann-2012} for 95\% of the pixels.

We regard the reconstruction presented here as an improvement over the reconstruction of \citet{oppermann-2012}. The results of \citet{oppermann-2012} should only be used in cases in which one explicitly does not want to be influenced by one of the assumptions we made here, like the spectral smoothness prior, the assumption that error variances should not be lower than quoted in the observational catalogs, or the explicit split of the data into the two categories.

\section{Summary}
\label{sec:conclusions}

We have studied the contributions to the observed Faraday rotation of extragalactic sources that are due to the Galactic interstellar medium, due to extragalactic magnetic fields, and due to observational noise. Extracting any of these three contributions is non-trivial, as they are superimposed on every line of sight. Another complication is that the observational error bars do not in every case describe the data likelihood accurately. This makes even a probabilistic analysis of the fractions of the data values due to the three different constituents challenging.

If the observations were noiseless, the extragalactic contributions could be estimated by simply subtracting an estimate of the Galactic foreground from the data values. However, in reality the observations are noisy and an estimate of the extragalactic contributions calculated in this way will contain this noise as well. Simply subtracting a Galactic foreground from the data is therefore not a good way of estimating extragalactic contributions. Furthermore, any estimate of the Galactic foreground will itself be uncertain and this uncertainty, when not taken properly into account, will introduce artifacts in the extragalactic estimate.

In our considerations, we strictly made the distinction between a physical quantity and an estimate of this quantity. The latter aims to equal the former, but, even if calculated correctly, there is always uncertainty involved and artifacts in the estimate may result. Taking into account the uncertainty of the estimate, however, should remove the artifacts. An example of such an artifact is the latitude dependence that we observed in the posterior mean estimates for the extragalactic Faraday depths. This latitude dependence vanishes once the uncertainty is taken into account.

To treat the complete problem of estimating the amount of both Galactic and extragalactic Faraday rotation from observations, we extended the algorithm of \citet{oppermann-2012}. This extended algorithm is based on a split of the data into a subset for which the observational error bars describe the data likelihood sufficiently and another subset for which this is not the case. It includes the estimation of the angular power spectrum of the Galactic foreground, assumed to be statistically isotropic up to a single latitude-dependent modulation, the estimation of this latitude-dependent function, the estimation of corrected noise variances for the subset of the data for which this is deemed necessary, and the estimation of the variance of the extragalactic contributions. We showed in a simulated scenario that all of these quantities are accurately reconstructed by our algorithm if our statistical model, including the split of the data, is correct.

For the application to observational data, we have considered several different ways to split the data into the two categories. We find that the most robust outcomes are achieved with splits that only regard a small fraction of the data (we use $1.75\%$ of the data points) situated near the Galactic poles as not afflicted by potential problems in the description of the data likelihood. In these cases we find extragalactic dispersions between $\SI{6.6}{\radian/m^2}$ and $\SI{7.2}{\radian/m^2}$. These numbers agree remarkably with the ones derived by \citet{schnitzeler-2010} by splitting the dispersion of observed Faraday rotation values into a latitude-dependent part, a contribution due to measurement errors, and a constant offset, deemed to be extragalactic in origin. Strictly speaking, both analyses only produce upper limits on the dispersion of the extragalactic contributions, but for slightly different reasons. While the estimate of \citet{schnitzeler-2010} may be increased due to a latitude-independent Galactic contribution, our estimate may be increased due to a Galactic contribution that is spatially uncorrelated on the scales probed by the observations.

We provide the derived estimates for all the involved quantities online at \url{http://www.mpa-garching.mpg.de/ift/faraday/}. We explain in Appendix~\ref{app:website} how these estimates can be used to estimate related quantities. The foreground products can be seen as updated versions of the results of \citet{oppermann-2012} that should be used preferentially, except in special cases where one of the assumptions we made in this paper is at question. We also provide 1\,000 samples of extragalactic contributions to the observed Faraday rotation, drawn from the posterior PDF for this quantity. This will enable future studies of extragalactic Faraday rotation to take into account the full probability distribution for these values, by performing any analysis on the set of samples rather than only on the posterior mean estimate. It should be noted that, within the framework of our assumptions, the extragalactic contributions are not very well constrained by the data. This is to some extent due to allowing the observational error bars of sources to get increased during the reconstruction, which increases the uncertainty of all reconstructed quantities. In addition, sources for which such an increase of the error bar can happen in our reconstruction will not have large estimates of the extragalactic contribution.

All our considerations point toward the importance of understanding the uncertainties of Faraday rotation measurements. For future surveys, this means that not only should the largest possible interval in $\lambda^2$-space be covered, but, as already pointed out by \citet{farnsworth-2011} and \citet{farnes-2014}, all the available information should be used in the data reduction, including the behavior of polarization fraction with frequency, as this can help avoid some of the rather poorly understood effects in RM synthesis studies that can lead to faulty estimates.

\begin{acknowledgements}
	The authors would like to thank the anonymous referee for constructive criticism, as well as Shane P.\ O'Sullivan for providing his unpublished rotation measure synthesis data, and S.\ Ann Mao for providing her newest data set in digital form.
	The results in this publication have been derived using the
        \texttt{NIFTy}\footnote{\url{http://www.mpa-garching.mpg.de/ift/nifty/}}
        package \citep{selig-2013}, as well as the
        \texttt{HEALPix}\footnote{\url{http://healpix.sf.net}} package
        \citep{gorski-2005}. This research has made use of NASA's
        Astrophysics Data System. It was supported by the DFG Forschergruppe 1254 ``Magnetisation of Interstellar and Intergalactic Media: The Prospects of Low-Frequency Radio Observations''. T.A.\ is supported by the Japan Society for the Promotion of Science (JSPS). B.M.G.\ acknowledges the support of the Australian Research Council through grant FL100100114. Some of the calculations were carried out on the Canada Foundation for Innovation funded CITA Sunnyvale cluster.
\end{acknowledgements}

\bibliographystyle{myaa}
\bibliography{note_final}

\begin{appendix}

\section{Reasons for imperfectly described likelihood functions}
\label{app:npi}

Usually, the likelihood for the observed Faraday depth of a source is assumed to be Gaussian and parameterized by an error bar $\sigma_i$, taken to be the standard deviation of this Gaussian. However, this may in some cases not be an accurate description of all unknown effects influencing the observations. For cases that cannot be described with sufficient precision by just this one parameter, the outcome of trying to describe them in this way can be misleading. For example, the likelihood could have a sharp peak that can be approximated as a narrow Gaussian, but significant sidelobes that are neglected in this description.

\begin{figure}
	\input{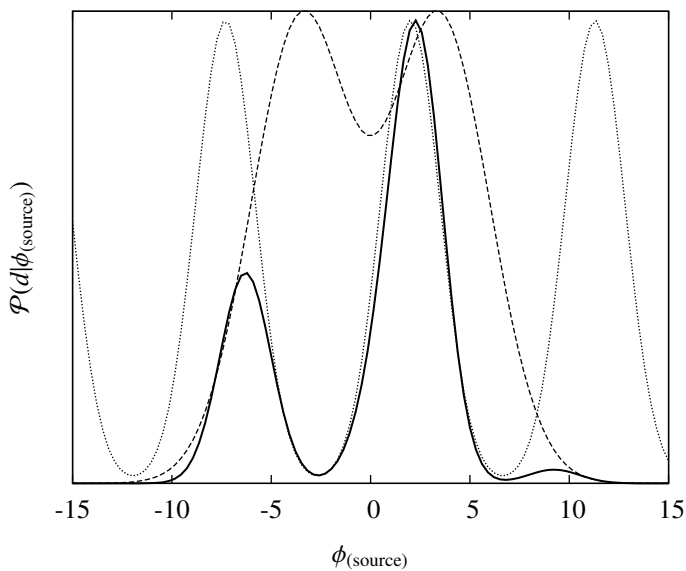}
	\caption{\label{fig:npi}Illustration of a possible likelihood curve for a single Faraday depth measurement. The dotted line shows the likelihood arising from the $n$-$\pi$ ambiguity of position angle measurements and the uncertainty in these measurements. The dashed curve shows an additional constraint on the magnitude of the source's Faraday depth and the solid curve is the resulting combined likelihood curve, i.e., the product of the dotted and dashed curves. The plotted curves have been arbitrarily rescaled to show similar amplitudes.}
\end{figure}

One possible cause for such a non-trivial likelihood function could be the inherent $n$-$\pi$ ambiguity of polarization orientation measurements. This is a problem for Faraday depth estimates from a linear $\lambda^2$-fit, but not for RM synthesis studies. We illustrate this in Fig.~\ref{fig:npi}. Assuming polarization angle measurements at a few frequencies and no other information still allows in principle for infinitely many discrete possible values for the source's Faraday depth. Any uncertainty in the measurement of the angles (say, Gaussian) will turn these discrete possibilities into a series of equally likely peaks in the likelihood for the source's Faraday depth, as shown by the dotted line in Fig.~\ref{fig:npi}. Additional data, like the degree of bandwidth depolarization \citep[see, e.g.,][]{sunstrum-2010} can lead to an additional constraint on the magnitude of the source's Faraday depth, but does not hold information on the sign of the Faraday depth. An example for such a constraint, formalized as a likelihood curve, is shown by the dashed line in Fig.~\ref{fig:npi}. The combined likelihood, given by the product of the likelihood for the position angle measurements and the likelihood for the depolarization measurement, is shown by the solid curve. Evidently, the result can be highly non-Gaussian. The error bar that is quoted as a measure of the observational uncertainty could in this case correspond to the width of a single peak arising from the observational uncertainty in the measurement of the polarization angles. If this error bar is then interpreted as describing a Gaussian likelihood function, this likelihood function includes only the main peak and neglects any secondary peaks such as the ones visible in Fig.~\ref{fig:npi}.

Usually, the error bar on an observational estimate of a source's Faraday depth is estimated as being inversely proportional to the signal to noise ratio of the polarized intensity observation, as well as the width of the frequency coverage in $\lambda^2$-space. This relation is based on linear Gaussian error propagation from the observations of the individual Stokes parameters to a polarization angle and to the slope of a straight line fit to the polarization angle as a function of $\lambda^2$, as shown by \citet{brentjens-2005} in their Appendix A. The result of this formula can be seen as an estimate for the width of the main peak shown in Fig.~\ref{fig:npi}. However, it does not allow for the presence of $n$-$\pi$ ambiguities and the ensuing non-Gaussianity shown by the solid line in Fig.~\ref{fig:npi}. Furthermore, the Gaussian approximation to the observational uncertainty of a derived polarization angle is not perfect, as pointed out by \citet{wardle-1974}, and the estimation of the polarimetric noise in the first place can also be erroneous. These are effects that can lead to a general underestimation of the widths of the likelihood peaks, i.e., $\sigma_i$, as was detected by \citet{stil-2011} for the RM catalog of \citet{taylor-2009}.

Furthermore, even though the sources that are used here are compact, it is not necessarily guaranteed that their emission as a function of Faraday depth is perfectly described by a single component. More complicated Faraday spectra due to, e.g., the internal structure of the source or differential Faraday rotation in the foreground within the telescope beam can lead to complicated effects on the observational estimates of Faraday depth, especially when a linear fit of polarization angle versus $\lambda^2$ is performed \citep[see][]{farnsworth-2011}. The usual formula used for the observational uncertainty described above implicitly assumes that the relationship between polarization angle and squared wavelength is linear. If polarized emission happens over an extended range of Faraday depths, the questions of what the Faraday depth of the source is and how large the intrinsic Faraday rotation of the source is become somewhat ill-defined. It is then not clear what the PDF for the noise should be if the noise is defined as the difference between the observed Faraday depth and two numbers characterizing the Galactic foreground contribution and the extragalactic contribution.

\section{Derivation of the filter formulas}
\label{app:derivation}

Here we discuss the filter formulas we use, their derivation, and the necessary approximations, following the strategy outlined in Sect.~\ref{sec:strategy}. Throughout, we assume that the covariance matrices have the structure described in Sect.~\ref{sec:covariances}.

\subsection{Estimating the Galactic contribution}

In order to estimate the Galactic contribution to Faraday rotation, we first have to estimate the dimensionless signal field $s = \frac{\phi_\mathrm{g}}{p}$, for which we calculate the mean over the PDF
\begin{equation}
	\label{eq:posterior-s}
	\mathcal{P}{\left( s \left| d, (\tilde{C}_\ell)_\ell = (\hat{\tilde{C}}_\ell)_\ell, (\tilde{\eta}_i)_i = (\hat{\tilde{\eta}}_i)_i, \tilde{\eta}_\mathrm{e} = \hat{\tilde{\eta}}_\mathrm{e} \right.\right)}.
\end{equation}
Using the zero-mean Gaussian priors for this signal field, the extragalactic contributions, as well as the noise contribution with covariances $S$, $E$, and $N$, respectively, this PDF is again a Gaussian with covariance
\begin{equation}
	\label{eq:defD}
	D = \left( S^{-1} + R^\dagger \left( N + E \right)^{-1} R \right)^{-1}
\end{equation}
and mean
\begin{equation}
	\label{eq:rec-m}
	m = D R^\dagger \left( N + E \right)^{-1} d.
\end{equation}
This $m$ therefore becomes our estimate for the dimensionless signal field and the diagonal of the matrix $D$ a measure for its pixel-wise uncertainty. The corresponding estimate for the Galactic contribution is obtained simply by multiplying with the Galactic latitude profile,
\begin{equation}
	\hat{\phi}_\mathrm{g} = pm,
\end{equation}
and its uncertainty accordingly as
\begin{equation}
	\mathrm{diag}{\left( D_{\phi_\mathrm{g}} \right)} = p^2 \mathrm{diag}{\left(D\right)}.
\end{equation}

Of course, the operators $D$ and $\left(N + E\right)$ depend on our estimates of the unknown quantities $(C_\ell)_\ell$, $(\eta_i)_i$, and $\eta_\mathrm{e}$ and necessitate that we estimate these in separate steps.

\subsection{Estimating the extragalactic contribution}

For the extragalactic contribution, we repeat the analysis done for the Galactic contribution and simply swap the roles of the Galactic and extragalactic contributions. We therefore find again a Gaussian posterior
\begin{equation}
	\label{eq:ext_post}
	\mathcal{P}{\left( \phi_\mathrm{e} \left| d, (\tilde{C}_\ell)_\ell = (\hat{\tilde{C}}_\ell)_\ell, (\tilde{\eta}_i)_i = (\hat{\tilde{\eta}}_i)_i, \tilde{\eta}_\mathrm{e} = \hat{\tilde{\eta}}_\mathrm{e} \right.\right)} = \mathcal{G}{\left(\left. \phi_\mathrm{e} - \hat{\phi}_\mathrm{e} \right| D_\mathrm{e} \right)},
\end{equation}
where the covariance is given by
\begin{equation}
	D_\mathrm{e} = \left( E^{-1} + \left( RSR^\dagger + N \right)^{-1} \right)^{-1}
\end{equation}
and the mean and our estimate by
\begin{equation}
	\label{eq:rec-e}
	\hat{\phi}_\mathrm{e} = D_\mathrm{e} \left( RSR^\dagger + N \right)^{-1} d = E \left( E + N \right)^{-1} \left( d - \hat{\phi}_\mathrm{g} \right).
\end{equation}

\subsection{Estimating the angular power spectrum of the dimensionless auxiliary field}

To estimate the angular power spectrum, we maximize the PDF
\begin{equation}
	\mathcal{P}{\left(\left.(\tilde{C}_\ell)_\ell \right| d, (\tilde{\eta}_i)_i = (\hat{\tilde{\eta}}_i)_i, \eta_\mathrm{e} = \hat{\eta}_\mathrm{e}\right)},
\end{equation}
where a tilde denotes a logarithmic quantity, i.e., $\tilde{C}_\ell = \log{\left(C_\ell\right)}$. This PDF is calculated straightforwardly by multiplying the Gaussian likelihood function
\begin{equation}
	\label{eq:app-likelihood}
	\mathcal{P}{\left( d \left| s, (\tilde{C}_\ell)_\ell, (\tilde{\eta}_i)_i = (\hat{\tilde{\eta}}_i)_i, \eta_\mathrm{e} = \hat{\eta}_\mathrm{e} \right.\right)} = \mathcal{G}{\left(d - Rs,E+N\right)}
\end{equation}
with the Gaussian signal prior and the prior for the angular power spectrum, given by Eqs.~\eqref{eq:IG-prior}-\eqref{eq:prior-C}, and marginalizing over $s$. The result is
\begin{align}
	& \mathcal{P}{\left(\left.(\tilde{C}_\ell)_\ell \right| d, (\tilde{\eta}_i)_i = (\hat{\tilde{\eta}}_i)_i, \eta_\mathrm{e} = \hat{\eta}_\mathrm{e}\right)} \nonumber\\
	& \propto \left|S\right|^{-1/2} \left|D\right|^{1/2} \left(\prod_\ell C_\ell^{-\alpha_\ell + 1} \mathrm{e}^{-\frac{q_\ell}{C_\ell}} \right) \nonumber\\
	& ~~~~\times \exp{\left\{-\frac{1}{2} \tilde{C}^\dagger T \tilde{C} + \frac{1}{2} d^\dagger \left(E + N\right)^{-1} R D R^\dagger \left(E + N\right)^{-1} d\right\}},
\end{align}
where we have dropped all factors that are independent of the angular power spectrum. Equating the derivative of this function with respect to $\tilde{C}_\ell$ with zero leads to the equation for our estimate of the angular power spectrum,
\begin{equation}
	\label{eq:rec-C}
	\hat{C}_\ell = \frac{q_\ell + \frac{1}{2}\mathrm{tr}{\left( \left( mm^\dagger + D \right) S_{(\ell)} \right)}}{\rho_\ell/2 + \alpha_\ell - 1 + \left(T\hat{\tilde{C}}\right)_\ell}.
\end{equation}
Here, $S_{(\ell)}$ denotes an operator that projects a field on the sphere onto its $\ell$-th multipole and $\rho_\ell = 2\ell + 1$ is the number of degrees of freedom of the $\ell$-th multipole.

\subsection{Estimating the noise variance correction factors}

Similar to the estimation of the angular power spectrum, we multiply the likelihood
\begin{equation}
	\mathcal{P}{\left( d \left| s, (\tilde{\eta}_i)_i, (\tilde{C}_\ell)_\ell = (\hat{\tilde{C}}_\ell)_\ell, \eta_\mathrm{e} = \hat{\eta}_\mathrm{e} \right.\right)} = \mathcal{G}{\left(d-Rs,N+E\right)}
\end{equation}
with the signal prior and the prior for the noise variance correction factors, given by Eq.~\eqref{eq:prior-eta}, and marginalize over $s$, resulting in
\begin{align}
	& \mathcal{P}{\left(\left.(\tilde{\eta}_i)_i \right| d, (\tilde{C}_\ell)_\ell = (\hat{\tilde{C}}_\ell)_\ell, \eta_\mathrm{e} = \hat{\eta}_\mathrm{e}\right)} \nonumber \\
	& \propto \left|\left(E + N\right)\right|^{-1/2} \left|D\right|^{1/2} \left( \prod_i \eta_i^{-\beta_i + 1} \exp{\left(-\frac{r_i}{\eta_i}\right)} \right) \nonumber\\
	& ~~~~\times \exp{\left\{\frac{1}{2} d^\dagger \left(E + N\right)^{-1} R D R^\dagger \left( E + N\right)^{-1} d - \frac{1}{2} d^\dagger \left(E + N\right)^{-1}d\right\}}.
\end{align}
We have again dropped all factors that are independent of $\eta_i$. After differentiating with respect to $\tilde{\eta}_i$ and equating to zero we find our estimate
\begin{equation}
	\label{eq:rec-eta_i}
    \hat{\eta}_i = \frac{r_i + \frac{1}{2}\left( \left(d -
        Rm\right)_i^2 + \left(RDR^\dagger\right)_{ii}
        \right)}{\beta_i - 1/2}
\end{equation}
for any data point in the SIP category.

\subsection{Estimating the extragalactic variance correction factor}

The calculation for the extragalactic variance correction factor is slightly more involved than the ones for the other estimators. We begin again by multiplying the likelihood
\begin{equation}
	\mathcal{P}{\left( d \left| s, (\tilde{\eta}_i)_i = (\hat{\tilde{\eta}}_i)_i, (\tilde{C}_\ell)_\ell = (\hat{\tilde{C}}_\ell)_\ell, \eta_\mathrm{e} \right.\right)}
\end{equation}
with the priors for the signal $s$ and for the variance correction factors, Eq.~\eqref{eq:prior-eta}. However, now we marginalize first over the error variance correction factors $(\eta_i)_i$. This leads to
\begin{align}
	& \mathcal{P}{\left(s, \log{\left(\eta_\mathrm{e}\right)} \left| d, (\tilde{C}_\ell)_\ell = (\hat{\tilde{C}}_\ell)_\ell \right. \right)} \nonumber\\
	& \propto \left( \prod_{j\in\mathrm{(VIP)}} \left(\eta_\mathrm{e} \sigma_\mathrm{e}^2 + \sigma_j^2 \right)^{-1/2}  \right) \eta_\mathrm{e}^{-\beta_\mathrm{e} + 1} \exp{\left(-\frac{r_\mathrm{e}}{\eta_\mathrm{e}}\right)} \nonumber\\
	& ~~~~\times \left( \prod_{i\in\mathrm{(SIP)}} \left( r_i + \frac{1}{2} \frac{(d-Rs)_i^2}{\sigma_i^2 + \sigma_\mathrm{e}^2} \right)^{1/2 - \beta_i} \right) \nonumber\\
	& ~~~~\times \exp{\left\{-\frac{1}{2} \left( d - Rs \right)^\dagger \left(E + N\right)_{\mathrm{(VIP)}}^{-1} \left( d - Rs \right)\right\}} \, \mathcal{G}{\left(s,S\right)},
\end{align}
where the first product is to be taken over all data points in the VIP category and the second product over all data points in the SIP category. $(E + N)_{\mathrm{(VIP)}}$ denotes the combination of the diagonal operator $(E + N)$ and the projection onto the data points of the VIP category. Here and in the following, we use the superscript ${-1}$ to denote the inverse for regular operators and the pseudo-inverse for singular operators, such as $(E + N)_{\mathrm{(VIP)}}$.

Marginalizing this PDF over the dimensionless signal field $s$ amounts to calculating the Gaussian integral
\begin{equation}
	\int\mathcal{D}s ~ \left( \prod_{i\in\mathrm{(SIP)}} \left( r_i + \frac{1}{2} \frac{(d-Rs)_i^2}{\sigma_i^2 + \sigma_\mathrm{e}^2} \right)^{1/2 - \beta_i} \right) \, \mathcal{G}{\left(s - m_{\mathrm{(VIP)}},D_{\mathrm{(VIP)}}\right)},
\end{equation}
where we have defined
\begin{equation}
	D_{\mathrm{(VIP)}} = \left( S^{-1} + R^\dagger \left(E + N\right)_{\mathrm{(VIP)}}^{-1} R\right)^{-1}
\end{equation}
and
\begin{equation}
	\label{eq:m_2}
	m_{\mathrm{(VIP)}} = D_{\mathrm{(VIP)}} R^\dagger \left( E + N \right)_{\mathrm{(VIP)}}^{-1} d.
\end{equation}
These would be the posterior covariance and mean for the dimensionless signal field if only the data points of the VIP category existed. Calculating this integral analytically for a positive value of $\beta_i$ is not possible. We therefore Taylor-expand the product in $\left(\frac{(d - Rs)_i^2}{\sigma_i^2 + \sigma_\mathrm{e}^2}\right)_i$ up to first order around its expectation value, given by
\begin{equation}
	\left< \frac{(d - Rs)_i^2}{\sigma_i^2 + \sigma_\mathrm{e}^2} \right>_{\mathcal{G}{(s - m_{\mathrm{(VIP)}},D_{\mathrm{(VIP)}})}} = \frac{\left(d - Rm_{\mathrm{(VIP)}}\right)_i^2 + \left(RD_{\mathrm{(VIP)}}R^\dagger\right)_{ii}}{\sigma_i^2 + \sigma_\mathrm{e}^2}.
\end{equation}
After the integration, the first order expansion term vanishes by definition and we are left with the zero-order term. Altogether, the PDF we are maximizing becomes
\begin{align}
	\label{eq:Poflogetae}
	 & \mathcal{P}{\left( \log{\left(\eta_\mathrm{e}\right)} \left| d, (\tilde{C}_\ell)_\ell = (\hat{\tilde{C}}_\ell)_\ell \right. \right)} \nonumber\\
	 & \propto \left|D_{\mathrm{(VIP)}}\right|^{1/2} \left( \prod_{j\in\mathrm{(VIP)}} \left( \eta_\mathrm{e} \sigma_\mathrm{e}^2 + \sigma_j^2 \right)^{-1/2} \right) \eta_\mathrm{e}^{-\beta_\mathrm{e} + 1} \mathrm{e}^{-\frac{r_\mathrm{e}}{\eta_\mathrm{e}}} \nonumber\\
	 & ~~~~\times \left( \prod_{i\in\mathrm{(SIP)}} \left( r_i + \frac{1}{2} \frac{\left( d - Rm_{\mathrm{(VIP)}} \right)_i^2 + \left( RD_{\mathrm{(VIP)}}R^\dagger \right)_{ii}}{\sigma_i^2 + \sigma_\mathrm{e}^2}\right)^{1/2 - \beta_i} \right) \nonumber \\
	& ~~~~\times \exp\Bigg\{-\frac{1}{2} d^\dagger \left(E + N\right)_{\mathrm{(VIP)}}^{-1} d \nonumber\\
	& ~~~~~~~~~~~~~~~~~+ \frac{1}{2} d^\dagger \left(E + N\right)_{\mathrm{(VIP)}}^{-1} R D_{\mathrm{(VIP)}} R^\dagger \left(E + N\right)_{\mathrm{(VIP)}}^{-1} d\Bigg\}.
\end{align}
The value of $\eta_\mathrm{e}$ that maximizes this function fulfills
\begin{equation}
	\label{eq:rec-eta_e}
	\hat{\eta}_\mathrm{e} = \frac{A + B}{C},
\end{equation}
where
\begin{equation}
	A = r_\mathrm{e} + \frac{\eta_\mathrm{e}^2 \sigma_\mathrm{e}^2}{2} \left( \sum_{j\in\mathrm{(VIP)}} \frac{\left(d - Rm_{\mathrm{(VIP)}}\right)_j^2 + \left( RD_{\mathrm{(VIP)}}R^\dagger \right)_{jj}}{\left( \eta_\mathrm{e} \sigma_\mathrm{e}^2 + \sigma_j^2 \right)^2} \right),
\end{equation}
\begin{align}
	B = & \sum_{i\in\mathrm{(SIP)}} \Bigg\{ \frac{\eta_\mathrm{e}^2\sigma_\mathrm{e}^2 \left(\beta_i - \frac{1}{2}\right)}{2} \nonumber\\
	& ~~\times \Bigg[ 2 \left(d - Rm_{\mathrm{(VIP)}}\right)_i \left( RD_{\mathrm{(VIP)}}R^\dagger \left(E + N\right)_{\mathrm{(VIP)}}^{-2} \left(d - Rm_{\mathrm{(VIP)}}\right) \right)_i \nonumber \\
	& ~~~~~~ - \left( RD_{\mathrm{(VIP)}}R^\dagger \left(E + N\right)_{\mathrm{(VIP)}}^{-2} RD_{\mathrm{(VIP)}}R^\dagger \right)_{ii} \Bigg] ~\Big/ \nonumber \\
	& ~~\left[ r_i \left( \sigma_i^2 + \sigma_\mathrm{e}^2 \right) + \frac{1}{2} \left(d - Rm_{\mathrm{(VIP)}}\right)_i^2 + \frac{1}{2} \left(RD_{\mathrm{(VIP)}}R^\dagger\right)_{ii} \right] \Bigg\},
\end{align}
and
\begin{equation}
	C = \beta_\mathrm{e} - 1 + \frac{1}{2} \left( \sum_{j\in\mathrm{(VIP)}} \frac{\eta_\mathrm{e} \sigma_\mathrm{e}^2}{\eta_\mathrm{e}\sigma_\mathrm{e}^2 + \sigma_j^2} \right).
\end{equation}

\subsection{Estimating the Galactic latitude profile}
\label{app:rec-p}

So far, we have assumed the Galactic latitude profile to be known. Another global iteration step will be needed to include this as a quantity to be reconstructed. In a first step, we calculate the profile function simply as the root mean square of all the data values in latitude bins. In doing this, we subtract the noise variance and smooth the squares with a Gaussian kernel of \SI{4}{\degree} full width at half maximum (FWHM). After the iteration of the filter equations derived here has converged using this profile function, we calculate an approximative map of the posterior mean for the squared Galactic Faraday depth according to
\begin{equation}
    \left< \phi_\mathrm{g}^2 \right>_{(\phi_\mathrm{g}|d)} \approx \left< (ps)^2 \right>_{\mathcal{G}{(s - m,D)}} = p^2 m^2 + p^2 \mathrm{diag}{(D)},
\end{equation}
where $p$ is the profile function used in the previous iteration steps. We then smooth this map again with a Gaussian kernel of \SI{4}{\degree} FWHM and average over Galactic latitude bins to obtain a new profile function. This procedure is repeated until the profile function has converged as well. As demonstrated in Sect.~\ref{sec:simulation}, a few of these global iteration steps suffice to achieve convergence.

\subsection{Implementation}

We start our reconstruction with a starting guess for the latitude-dependent profile function and the angular power spectrum. The initial profile is calculated directly from the data as described in Sect.~\ref{sec:strategy}. For the angular power spectrum, we choose as a starting guess a simple power law,
\begin{equation}
	C_\ell = 1.53 \, \ell^{-2.17},
\end{equation}
based on the results of \citet{oppermann-2012}.
We then iterate the following steps until convergence:
\begin{itemize}
	\item Calculate a new estimate of the Galactic Faraday depth according to Eq.~\eqref{eq:rec-m}.
	\item As an auxiliary field, calculate the estimate of the Galactic Faraday depth using only the data points of the VIP category, $m_{\mathrm{(VIP)}}$, according to Eq.~\eqref{eq:m_2}.
	\item Use these current estimates to update the estimates for the error variance correction factors $(\eta_i)_i$ and the correction factor for the extragalactic variance $\eta_\mathrm{e}$ according to Eqs.~\eqref{eq:rec-eta_i} and \eqref{eq:rec-eta_e}.
	\item Use these to update the estimate of the angular power spectrum according to Eq.~\eqref{eq:rec-C}.
\end{itemize}
After this iteration has converged, we calculate a new profile function as described in Appendix~\ref{app:rec-p} and repeat the whole procedure until the profile function has converged as well. At any point, the estimate for the extragalactic contributions can be calculated via Eq.~\eqref{eq:rec-e}.

Due to the high dimensionality of the involved vector spaces (41\,330 data points in the simulation, 41\,632 observational data points, 196\,608 pixels in our maps), we avoid treating the involved operators as explicit matrices. Operator inversions are performed with a conjugate gradient routine and whenever diagonal elements are needed explicitly, these are estimated via the technique of operator probing as implemented in the \texttt{NIFTy} package \citep{selig-2013}. These methods yield only approximate solutions that can lead to artifacts in the results, such as the increased posterior variance discussed in Appendix~\ref{app:howtousesamples}.

\section{Additional considerations for observed extragalactic Faraday rotation}
\label{app:bryan}

As noted in Sects.~\ref{sec:illustration} and \ref{sec:firstsixsplits}, the difference between the observed Faraday
depth, $d_i$, and the estimated Galactic contribution, $\hat{\phi}_{\mathrm{g},i}$,
is split between the estimates of the extragalactic contribution,
$\hat{\phi}_{\mathrm{e},i}$, and the noise contribution, $\hat{n}_i$, as per Eqs.~\eqref{eq:split1} and \eqref{eq:split2}.

This choice then has the added consequence that the inequalities
\begin{equation}
	\label{eq:bias}
	\hat{\phi}_{\mathrm{e},i} \le | d_i - \hat{\phi}_{\mathrm{g},i} | \\
	n_i \le | d_i - \hat{\phi}_{\mathrm{g},i} |
\end{equation}
always hold, i.e., that the most probable estimates of the extragalactic
contribution and of the noise always have the same sign. The underlying
physical cause for these inequalities is our explicit adoption of a Gaussian
PDF of zero mean for $\phi_\mathrm{e}$, as set out in Sect.~\ref{sec:covariances}. For a PDF of this form,
smaller intrinsic values of $|\phi_\mathrm{e}|$ are always more probable than larger
ones. Under such circumstances, $| d_i - \hat{\phi}_{\mathrm{g},i} |$ is most likely
an overestimate of $|\phi_{\mathrm{e},i}|$, a situation broadly analogous to the
Eddington bias seen in flux measurements
\citep{eddington-1913, eddington-1940}. Our approach corrects for
this bias, resulting in the inequalities in Eq.~\eqref{eq:bias} above.
We reiterate that $\hat{\phi}_{\mathrm{e},i}$ is only the most probable estimate of
${\phi}_{\mathrm{e},i}$, and that the true underlying value $|\phi_{\mathrm{e},i}|$ may be larger
than $| d_i - \hat{\phi}_{\mathrm{g},i} |$.

The choice of a Gaussian PDF of zero mean for $\phi_\mathrm{e}$ is physically
justified in the absence of further information, since this is the form
expected to result from a homogeneous randomly oriented population of
emitters and extragalactic intervenors. However, we note that further
observational work is still needed to improve the experimental underpinning
for the PDF adopted for $\phi_\mathrm{e}$. For example, throughout this paper, we
have implicitly assumed that the PDF for $\phi_\mathrm{e}$ is uncorrelated with the
signal-to-noise ratio of the measurements, i.e., that the extragalactic
Faraday depths of bright polarized radio sources will be drawn from the same
underlying probability distribution as for the extragalactic Faraday depths
of faint sources. However, effects such as depolarization, Doppler boosting,
redshift-dilution, and the overall cosmic evolution of gas densities and
magnetic field strengths can all couple the PDF for $\phi_\mathrm{e}$ to the
brightness of polarized sources in complicated ways.
In such cases, the PDF for $|\phi_\mathrm{e}|$ can potentially take on a positive
slope for some values of $|\phi_\mathrm{e}|$  or can even become multimodal, which
for small values of $n$ can negate or even reverse the inequalities in
Eq.~\eqref{eq:bias} above. We expect such cases to be rare, and their
effects on $\hat{\phi}_\mathrm{e}$ to be small, but such possibilities are worthy of
further investigation in future work.

\section{Online access and usage of the results}
\label{app:website}

At \url{http://www.mpa-garching.mpg.de/ift/faraday/} we provide the results of our study in the `polar caps' data split described in Sect.~\ref{sec:realworld}. All results are provided in binary format both as \texttt{hdf5} files and \texttt{fits} files. For the Galactic foreground, we provide maps of the reconstructed dimensionless signal field, $m$, the reconstructed Faraday depth, $\hat{\phi}_\mathrm{g}$, their uncertainty maps, as well as the profile function connecting the two and the angular power spectrum. For the extragalactic contributions, we provide $1\,000$ samples drawn from the Gaussian approximation to their posterior probability distribution, given by Eq.~\eqref{eq:posteriorforextragal}. In the following, we discuss their usage.

\subsection{How to use the posterior samples}
\label{app:howtousesamples}

\begin{figure}
	\input{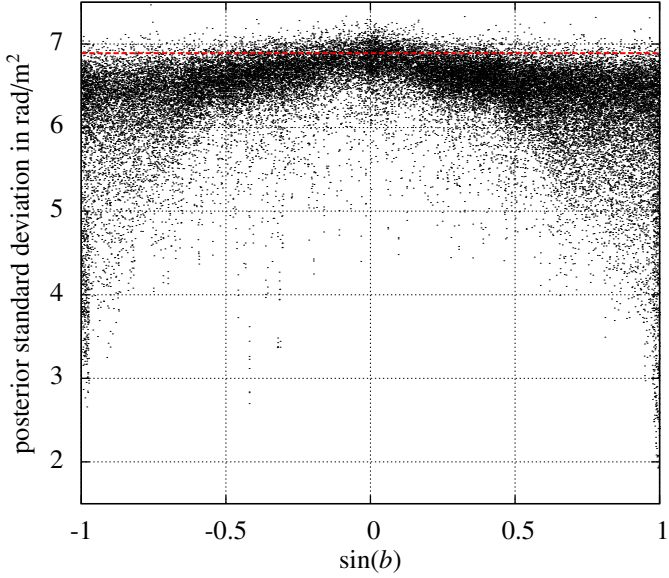}
	\caption{\label{fig:posteriorstddev}Standard deviation of the posterior for the extragalactic contribution to each data point versus $\sin(b)$. Plotted is the estimate calculated from the samples as the square root of Eq.~\eqref{eq:posteriorstddev}. The horizontal dashed line shows the prior standard deviation of $\sigma_\mathrm{e} = \SI{6.9}{\radian/m^2}$.}
\end{figure}

\begin{figure}
	\input{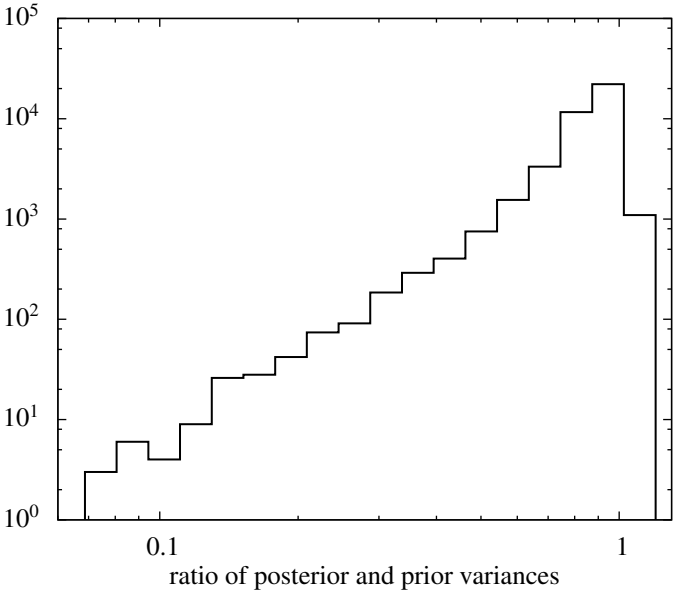}
	\caption{\label{fig:information_gain}Histogram of the ratios of the posterior variance for the extragalactic contribution to each data point, as calculated from the samples via Eq.~\eqref{eq:posteriorstddev}, to the prior variance. Both axes are scaled logarithmically.}
\end{figure}

A range of values that is less likely appears less often in the samples and vice versa. Thus, the frequency with which the sample values lie within a certain interval gives the posterior probability for the true extragalactic contribution to lie within that interval.

When calculating a quantity as a function of the extragalactic Faraday contribution for one or several sources,
\begin{equation}
	f{\left( \phi_{\mathrm{e},1}, \phi_{\mathrm{e},2}, \dots, \phi_{\mathrm{e},41\,632} \right)},
\end{equation}
this function should be evaluated for each of the samples. This will yield 1\,000 different answers,
\begin{equation}
	f^{(k)} = f{\left( \phi_{\mathrm{e},1}^{(k)}, \phi_{\mathrm{e},2}^{(k)}, \dots, \phi_{\mathrm{e},41\,632}^{(k)} \right)}, ~~~ k = 0,\dots,999,
\end{equation}
where $\phi_{\mathrm{e},i}^{(k)}$ is the value for the extragalactic contribution to the $i$-th source according to the $k$-th sample. In the limit of infinitely many samples, the distribution of these answers gives the posterior distribution for the quantity of interest $f$, given the data and assumptions that we have used and the approximations that we have made. In practice, a finite number of samples has to be used. The more samples are used, the more accurate the resulting distribution.

Finally the probability density for $f$ approximated thusly can again be summarized, e.g., by calculating its mean
\begin{equation}
	\left< f \right>_{(f|d)} \approx \frac{1}{1000} \sum_{k=0}^{999} f^{(k)}
\end{equation}
and its (co)variance
\begin{align}
	&\left< \left(f - \left< f \right>_{(f|d)}\right) \left(f - \left< f \right>_{(f|d)}\right)^\dagger \right>_{(f|d)} \nonumber\\
	& \approx \frac{1}{1000} \sum_{k=0}^{999} \left( f^{(k)} - \frac{1}{1000} \sum_{k'=0}^{999} f^{(k')} \right) \left( f^{(k)} - \frac{1}{1000} \sum_{k'=0}^{999} f^{(k')} \right)^\dagger.
\end{align}
These formulas are equally true for scalar functions $f$ and vector-valued functions $f$.

Thus, we can for example calculate the posterior mean for the extragalactic contribution to the $i$-th data point as
\begin{equation}
	\left< \phi_{\mathrm{e},i} \right>_{(\phi_{\mathrm{e},i}|d)} \approx \frac{1}{1000} \sum_{k=0}^{999} \phi_{\mathrm{e},i}^{(k)}
\end{equation}
and the posterior variance for the $i$-th data point as
\begin{equation}
	\label{eq:posteriorstddev}
	\left< \left(\phi_{\mathrm{e},i} - \left<\phi_{\mathrm{e},i}\right>_{(\phi_{\mathrm{e},i}|d)} \right)^2 \right>_{(\phi_{\mathrm{e},i}|d)} \approx \frac{1}{1000} \sum_{k=0}^{999} \left( \phi_{\mathrm{e},i}^{(k)} - \frac{1}{1000} \sum_{k'=0}^{999} \phi_{\mathrm{e},i}^{(k')} \right)^2.
\end{equation}
In the last two columns of the provided files, we give this mean and the standard deviation, i.e., the square root of the last expression.

The posterior mean is also plotted in the top panel of Fig.~\ref{fig:exoflat_polarcaps} and we show the posterior standard deviations in Fig.~\ref{fig:posteriorstddev}. The approximate posterior we use in the calculation of the posterior mean estimate and in the drawing of the samples corresponds to a Gaussian posterior after fixing the prior covariances for the Galactic and extragalactic contributions and for the noise, i.e., to Eq.~\eqref{eq:posteriorforextragal}. Consequently, the uncertainty due to the uncertain reconstruction of the angular power spectrum $(C_\ell)_\ell$, the error variance correction factors $(\eta_i)_i$, and the correction factor for the extragalactic variance $\eta_\mathrm{e}$, is no longer included in this PDF. One logical consequence is that the posterior standard deviations for the extragalactic contributions, which give a measure of our uncertainty after considering the data, should in every case be smaller than the corresponding prior standard deviation, which we have reconstructed to be $\hat{\sigma}_\mathrm{e}=\SI{6.9}{\radian/m^2}$ in the `polar caps' split that is used here. Figure~\ref{fig:posteriorstddev} shows that this is not strictly the case for some of the data points. This is due to the approximate nature of the sampling procedure and the use of approximate iterative schemes for matrix inversion. The posterior standard deviations plotted in Fig.~\ref{fig:posteriorstddev} behave as expected, being in general slightly lower than the prior standard deviation, and more so nearer to the poles, where the sensitivity to the extragalactic contributions is largest.

In Fig.~\ref{fig:information_gain}, we plot a histogram of the ratio of the posterior variance as estimated from the samples via Eq.~\eqref{eq:posteriorstddev} and the prior variance $\hat{\sigma}_\mathrm{e}^2$. This ratio can be roughly interpreted as a measure for the constraining power of the data, since it compares the uncertainty after considering the data to the uncertainty before. We note, however, that in our reconstruction, the prior variance was itself reconstructed from the data, so we have actually extracted more information from the data. A smaller ratio in Fig.~\ref{fig:information_gain} means more constraining power, with a ratio of 1 meaning no new constraint at all. As we explained in the previous paragraph, a ratio larger than 1 is not allowed mathematically. Figure~\ref{fig:information_gain} shows that the ratio is close to 1 for most of the sources, meaning that the data do not constrain the extragalactic contribution to an individual source much.

\subsection{Correlations}

\begin{table}
	\centering
	\caption{\label{tab:correlation_sources}Specifications of the four sources for which sample values are plotted in Fig.~\ref{fig:posteriorcorrelations}.}
	\begin{tabular}{l c l r @{.}l r @{.}l}
		\hline
		\hline
		panel & source & catalog & \multicolumn{2}{c}{$l/\SI{}{\degree}$} & \multicolumn{2}{c}{$b/\SI{}{\degree}$} \\
		\hline
		top & A & Taylor & -3 & 5608535 & -5 & 6028647\\
		top & B & Taylor & -3 & 5593824 & -5 & 5575824\\
		bottom & C & O'Sullivan & -45 & 865494 & -27 & 910206\\
		bottom & D & O'Sullivan & -45 & 973431 & -28 & 065784\\
		\hline
	\end{tabular}
\end{table}

\begin{figure}
	\input{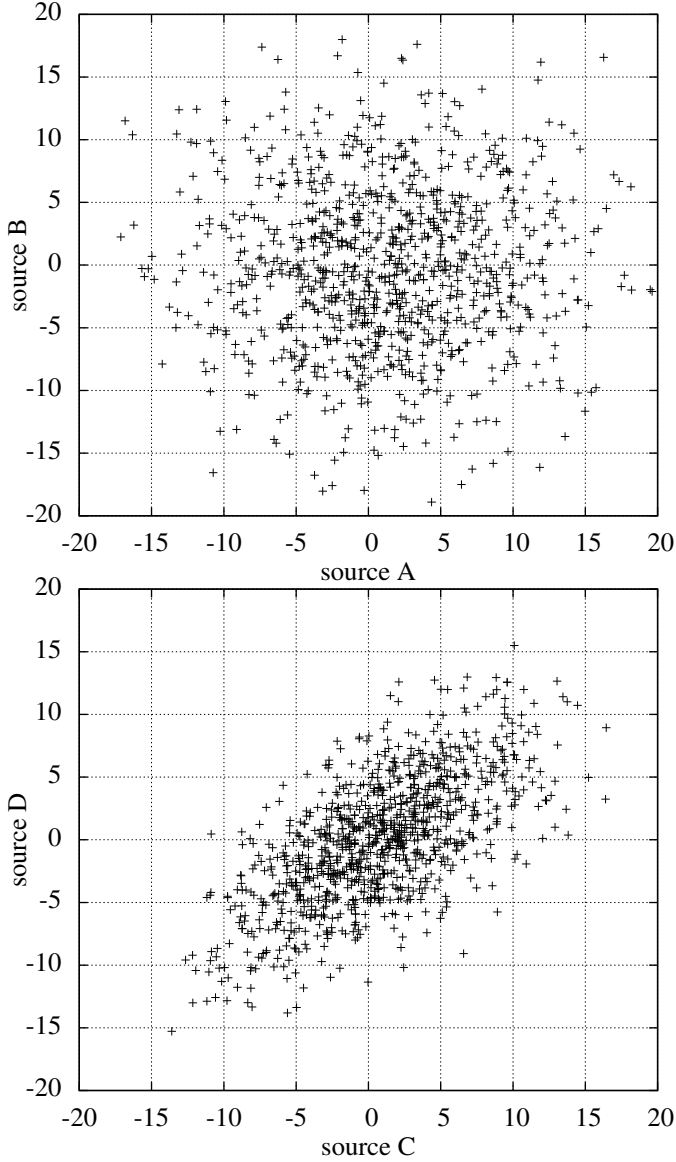}
	\caption{\label{fig:posteriorcorrelations}Posterior sample values for the extragalactic contributions for the four sources listed in Table~\ref{tab:correlation_sources}. All axes are in \SI{}{\radian/m^2}.}
\end{figure}

It should also be noted that correlations are present in the posterior distribution for the extragalactic contributions. To illustrate this, we plot sample values for two pairs of sources that are nearby one another in Fig.~\ref{fig:posteriorcorrelations}. The sources are described in Table~\ref{tab:correlation_sources}. Although both panels of the figure show a pair of sources that is very close, only one of them shows significant correlations. In conclusion, for some sources the posterior uncertainty of the extragalactic contributions is strongly correlated, for others not. This complicated correlation structure is automatically included when an analysis is performed using the samples as described in this appendix but is lost completely if only posterior mean and variance for each source individually are considered.

\subsection{Change of prior}

The samples discussed here describe the posterior PDF for the extragalactic contributions, which depends on the prior we have used for these, i.e., the uncorrelated Gaussian distribution with a standard deviation of $\hat{\sigma}_\mathrm{e} = \SI{6.9}{\radian/m^2}$. We will denote this prior as $\mathcal{P}{(\phi_\mathrm{e}|\hat{\sigma}_\mathrm{e} = \SI{6.9}{\radian/m^2})}$ in the following. However, the samples can even be used to calculate expectation values of a function $f$ with respect to a posterior distribution $\mathcal{P}{(\phi_\mathrm{e}|d,X)}$ that is based on a new prior $\mathcal{P}{(\phi_\mathrm{e}|X)}$. This can be seen from a simple application of Bayes' theorem.

We can write the expectation value with respect to the new posterior as
\begin{align}
	&\int \mathcal{D}\phi_\mathrm{e} \, f{(\phi_\mathrm{e})} \, \mathcal{P}{(\phi_\mathrm{e}|d,X)}\nonumber\\
	&\propto \int \mathcal{D}\phi_\mathrm{e} \, f{(\phi_\mathrm{e})} \, \mathcal{P}{(d|\phi_\mathrm{e})} \, \mathcal{P}{(\phi_\mathrm{e}|X)} \nonumber \\
	&= \int \mathcal{D}\phi_\mathrm{e} \, f{(\phi_\mathrm{e})} \, \mathcal{P}{(d|\phi_\mathrm{e})} \, \mathcal{P}{(\phi_\mathrm{e}|\hat{\sigma}_\mathrm{e} = \SI{6.9}{\radian/m^2})} \nonumber\\
	&~~~~~~~\times \, \frac{\mathcal{P}{(\phi_\mathrm{e}|X)}}{\mathcal{P}{(\phi_\mathrm{e}|\hat{\sigma}_\mathrm{e} = \SI{6.9}{\radian/m^2})}} \nonumber \\
	& = \int \mathcal{D}\phi_\mathrm{e} \, f{(\phi_\mathrm{e})} \, \mathcal{P}{(\phi_\mathrm{e}|d,\hat{\sigma}_\mathrm{e} = \SI{6.9}{\radian/m^2})} \, \frac{\mathcal{P}{(\phi_\mathrm{e}|X)}}{\mathcal{P}{(\phi_\mathrm{e}|\hat{\sigma}_\mathrm{e} = \SI{6.9}{\radian/m^2})}},
\end{align}
so as an expectation value of the function $f{(\phi_\mathrm{e})} \, \frac{\mathcal{P}{(\phi_\mathrm{e}|X)}}{\mathcal{P}{(\phi_\mathrm{e}|\hat{\sigma}_\mathrm{e} = \SI{6.9}{\radian/m^2})}}$ with respect to the original posterior. In practice this means that one has to calculate a weighted average of the function $f$ evaluated on the samples,
\begin{equation}
	\int \mathcal{D}\phi_\mathrm{e} \, f{(\phi_\mathrm{e})} \, \mathcal{P}{(\phi_\mathrm{e}|d,X)} \approx \frac{1}{W} \sum_{k=0}^{999} f^{(k)} \, w^{(k)},
\end{equation}
where the weights are given by the prior ratios
\begin{equation}
	w^{(k)} = \frac{\mathcal{P}{(\phi_\mathrm{e}^{(k)}|X)}}{\mathcal{P}{(\phi_\mathrm{e}^{(k)}|\hat{\sigma}_\mathrm{e} = \SI{6.9}{\radian/m^2})}}
\end{equation}
and
\begin{equation}
	W = \sum_{k=0}^{999} w^{(k)}.
\end{equation}

\end{appendix}

\end{document}